\documentclass[journal]{IEEEtran}

\usepackage{graphicx,amsmath,amssymb,overpic,here,multirow}          

\newtheorem{corollary}{\bf Corollary}
\newtheorem{propos}{\bf Proposition}[section]
\newtheorem{lemma}{\bf Lemma}[section]
\newtheorem{definition}{\bf Definition}[section]

\newtheorem{remark}{\bf Remark}[section]
\newtheorem{assumption}{\bf Assumption}[section]

\usepackage[dvipsnames]{xcolor}
\usepackage{tikz}
\usepackage{pgfplots}
\usetikzlibrary{positioning}

\usepackage{algorithm}
\usepackage{algpseudocode}

\ifCLASSINFOpdf
\else
\fi
\begin{document}
%
\title{A State-Dependent Updating Period For \\ Certified Real-Time Model Predictive Control}
%
%
%

\author{Mazen Alamir
\thanks{M. Alamir is with CNRS/Gipsa-lab, Control Systems Dept., University of Grenoble, France e-mail: mazen.alamir@grenoble-inp.fr  (http://www.mazenalamir.fr).}
}

\maketitle

\begin{abstract}
In this paper, a state-dependent control updating period framework is proposed that leads to real-time implementable Model Predictive Control with certified practical stability results and constraints satisfaction. The scheme is illustrated and validated using new certification bound that is derived in the case where the Fast Gradient iteration is used through a penalty method to solve generally constrained convex optimization problems. Both the certification bound computation and its use in the state-dependent updating period framework are illustrated in the particular case of linear MPC. An illustrative example involving a chain of four integrators is used to show the explicit computation of the state-dependent control updating scheme. 
\end{abstract}

\begin{IEEEkeywords}
MPC, Certification, Real-time, Stability.
\end{IEEEkeywords}

%
\IEEEpeerreviewmaketitle

\section{Introduction}
Modern control paradigms such as Model Predictive Control \cite{Mayne:2000}, Moving-Horizon Observers \cite{IJCobs:1999}  or adaptive identification of varying models \cite{Zhiwen:2000}  to cite but few issues involve the real-time, on-line solution of constrained optimization problems. In such applications, the output of the optimizer (namely the sub-optimal solution of the optimization problem) is fed to some neighboring modules in order to achieve some engineering tasks. The quality of the global task  may strongly depend on the quality of the sub-optimal solution and the frequency with which it can be updated by the optimizer and since this solution has to be delivered in finite and probably short time, it is important to be able to precisely link the  quality of the suboptimal solution to the available computation time for a predefined embedded computation power. When the latter is not yet defined, such insight enables to choose the appropriate computational power given the required quality of the sub-optimal solution. \ \\ \ \\ 
The last few years witnessed an increasing interest in the certification issue \cite{Richter:2012,Jones2012,Bemporad2012}. These almost simultaneous works proposed certification bounds for fast gradient-based iterations \cite{Nesterov2004,Nesterov1983} applied to Quadratic Programming (QP) problems involving only simple constraints that enable easy projection on the admissible set. Otherwise, the iterations that are needed to perform the projection have to be counted as well and certified with some associated lower bounds which would invalidate the relevance of the proposed bounds. \ \\ \ \\ It is not surprising that recent certification-related results concerned fast gradient-based iterations. This is because the simplicity of this iteration and the associated low computational cost have been rapidly identified as appealing properties in the real-time context which is the very reason for which certification results were required. \ \\ \ \\ Regarding the other alternatives, active set iterations \cite{Ferreau:2008}, while computationally efficient and while showing a provably finite number of iterations to converge (for QP problems), seem to resist to the derivation of convergence rates which makes impossible the computation of certification bounds. As for interior point methods \cite{Domahidi2012o,Zavala:2008,Biegler:2010}, certification bounds exist \cite{McGovern:2000} but seem to be systematically over pessimistic \cite{Richter:2012}. Nevertheless, for many problems, it might still be more appropriate to use these efficient although uncertified or pessimistically certified algorithms rather than to use a slow certified iterations. The {\em right} choice is problem-dependent. \ \\ \ \\ 
{\bf The first part of this paper} belongs to the family of works that address the derivation of certification bounds for fast gradient-based iterations in the presence of general constraints. This is motivated by the nice properties mentioned above, namely the reduced complexity of the single associated iteration that enables the use of extremely short updating period. As it has been recently shown \cite{alamirecc13,alamirecc14,Bonne:2014arxiv}, this last property may compensate the drawback of  potentially higher number of iterations when compared to some alternative methods, especially in uncertain context (which includes perfectly known systems under unpredictable set-point dynamics). In such situations, as underlined by \cite{Nesterov2004}, it is important to distinguish between the concepts of {\em analytical complexity} which involves only the number of iterations (regardless of their inherent computational cost) and the {\em arithmetical complexity} which accounts for the total number of elementary operations until convergence which is obviously the appropriate indicator in real-time context and this is precisely why fast gradient is an interesting option. \ \\ \ \\ 
{\bf The second part of the paper} proposes a general framework to explicitly account for the arithmetical complexity by including the computation time for a single iteration in the overall convergence analysis and trade-off handling. This feature if absent from recent works on the certification issue such as \cite{Richter:2012} where the number of iteration is induced from the required precision on the solution and the corresponding number of iteration is derived as a consequence. This argumentation suggests that provided that one uses sufficiently high number of iterations, convergence of the real-time MPC will be guaranteed. This paper shows that this argument is generically erroneous and that in realistic situations, the appropriate updating period shows lower and upper bounds beyond which stability can no more be guaranteed.   \ \\ \ \\ 
More precisely, the contribution of the present paper lies in the following items: \ \\ \ \\ 
1) it gives a certification bound for the fast gradient algorithm when applied to solve a general (not necessarily simple bounded) convex optimization problems by means of a penalty approach. The number of iterations needed to achieve a prescribed level of precision on the optimal cost and a prescribed level of precision on the satisfaction of the soft constraints (while the hard constraints are fully satisfied) is given as a function of the problem's characteristics. \ \\ \ \\  
2) it shows how the certification bound so obtained can be used in the framework of real-time MPC in order to assess the practical asymptotic stability of the closed-loop performance under a state-dependent control updating period. The latter is computed based on some key properties of the MPC formulation. This second part while using the results of the first part has a general scope and can be applied to any available algorithm with computable certification bound. \ \\ \ \\ 
This paper is organized as follows: Section \ref{sec-pbstat} defines the class of optimization problems addressed in the paper together with the associated definitions and notation. Section \ref{secdnwa} states the working assumptions and gives some preliminary results that are used in the next sections. The algorithm and the associated certification bounds are presented in section \ref{secalgo} with instantiation to the specific case of QP problems. The use of the certification bounds in real-time MPC implementation through state-dependent control updating period is proposed in section \ref{apprtMPC} and the concrete computation of the parameters involved in the expressions is shown for the specific case of linear MPC. Finally, the whole scheme is illustrated through the MPC-based tracking problem for a quadruple integrators under state and control constraints. For the sake of clarity, all the technical proofs are gathered in appendix \ref{secappend} except those that can be given in few words.  

\section{Problem Statement} \label{sec-pbstat} 
Consider the following optimization problem in the decision variable $p$:
\begin{eqnarray}
\min_{p\in \mathbb{R}^{n_p}} f_0(p)\quad \vert \ \mbox{\rm $c_i(p)\le 0$ \ $\forall i\in I_h\cup I_s:=\{1,\dots,n_c\}$} \label{defproblem} 
\end{eqnarray} 
where $I_s$ and $I_h$ are the disjoint subsets of $\{1,\dots,n_c\}$ that define a partition of the set of constraints into soft and hard constraints respectively.  $f_0(\cdot)$ is the cost to be minimized while $c_i: \mathbb{R}^{n_p}\rightarrow \mathbb{R}$ defines the $i$-th inequality constraint. Note that saturation constraints on $p$ are supposed to be included in the set of inequality constraints. It is assumed that $f_0$ and $c_i$ are differentiable for all $i$. \ \\ \ \\ 
The algorithm proposed in this paper invokes the following penalty induced augmented cost:
\begin{eqnarray}
f(p):=f_0(p)+\rho\times\psi(p) \label{defdeextendedf} 
\end{eqnarray} 
where $\rho$ is called the penalty parameter while $\psi: \mathbb{R}^{n_p}\rightarrow \mathbb{R}_+$ is the constraints induced cost given by:
\begin{eqnarray*}
\psi(p):=\sum_{i\in I_s}\left[\max\{0,c_i(p)\}\right]^2 +\sum_{i\in I_h}\left[\max\{0,c_i(p)+\varepsilon_\psi\}\right]^2\label{defdepsi} 
\end{eqnarray*} 

For a given paire $\bar\varepsilon:=(\varepsilon_0,\varepsilon_\psi)$ of strictly positive reals, a candidate value $p$ is called an $\bar\varepsilon$-suboptimal solution of (\ref{defproblem}) if the following two conditions hold:
\begin{eqnarray}
\vert f_0(p)-f^{opt}\vert \le \varepsilon_0 \quad \mbox{\rm and}\quad \psi(p)\le \varepsilon_\psi^2 \label{bbHGFF} 
\end{eqnarray} 
where $f^{opt}$ denotes the optimal value of (\ref{defproblem}). \ \\ \ \\ 
The relevance of the second constraint in (\ref{bbHGFF}) lies in the fact that when satisfied, this constraint implies that all the hard constraints are rigorously satisfied while the maximum violation of any soft constraint is lower than $\varepsilon_\psi$.\ \\ \ \\ 
The first aim of the present paper is to derive the necessary relations that enable for a given precision $\bar\varepsilon$ to choose the appropriate penalty coefficient $\rho$ and the stopping condition for the fast gradient iteration to be used in the unconstrained minimization of the cost function $f$ defined by (\ref{defdeextendedf}). Moreover, the bound on the minimum number of iterations that guarantees an $\bar\varepsilon$-suboptimal solution to the original problem is derived. This is done in sections \ref{secdnwa} and \ref{secalgo}.  \ \\ \ \\ The second aim is to show that this certification result (or any similar one for possibly another algorithm) can then be used to design a real-time constrained MPC implementation in which a state-dependent control updating period is used to yield certified convergence properties. This is done in section \ref{apprtMPC}. \ \\ \ \\ 
The results are proved in a rather general convex settings and for both goals, the expressions enabling the parameters involved in the statements of the results to be computed are explicitly given in the specific case of QP problems and linear MPC design. 
\section{Assumptions and preliminary results} \label{secdnwa} 
\subsection{Definitions and Notation}
In what follows, $f^{'}(p)$, $f_0^{'}(p)$ and $\psi^{'}(p)$ denote the gradients of the functions w.r.t the decision variable $p$. The euclidien  norm of $f^{'}(p)$ is denoted by $g(p)=\|f^{'}(p)\|$. For a scalar continuously differentiable function $\ell$ defined on $\mathbb R^n$, the notation $\ell\in \mathcal S_\mu^1$  states that $\ell$ is a $\mu$-strongly convex function, namely for all $(p_1,p_2)$:
\begin{eqnarray}
\ell(p_2)\ge \ell(p_1)+\langle\ell^{'}(p_1),p_2-p_1\rangle+\dfrac{\mu}{2}\|p_2-p_1\|^2 \label{defdemu} 
\end{eqnarray} 
where $\mu$ is called the convexity parameter of $\ell$ \cite{Nesterov2004}. Similarly, the notation $\ell\in \mathcal F_L^1$  indicates that the continuously differentiable function $\ell$ satisfies for all $(p_1,p_2)$:
\begin{eqnarray}
\ell(p_2)\le \ell(p_1)+\langle\ell^{'}(p_1),p_2-p_1\rangle+\dfrac{L}{2}\|p_2-p_1\|^2 \label{defdeL} 
\end{eqnarray} 
When $\ell$ satisfies both (\ref{defdemu})-(\ref{defdeL}), the notation $\ell \in \mathcal S_{\mu,L}^1$ is used. The set $\mathcal C$ denotes the set of singular points of $f(\cdot)$, namely the set of $p$ such that $g(p)=0$. Given a subset $\mathcal A\subset \mathbb{R}^{n_p}$, the notation $d(p,\mathcal A)$ refers to the distance from $p$ to $\mathcal A$, namely $d(p,\mathcal A):=\min_{z\in \mathcal A}\|z-p\|$. The short notation $d(p):=d(p,\mathcal C)$ is used for the specific set $\mathcal C$. The set $\mathcal A_{\psi=0}$ is the set of $p$ such that $\psi(p)=0$. Given a bounded subset $\mathbb P$, $\delta_\mathbb{P}$ denote the radius of $\mathbb P$ namely $\delta_\mathbb{P}:=\sup_{(x_1,x_2)\in \mathbb P^2}\|x_1-x_2\|$. For a compact set $\mathbb X$, the notation $\varrho(\mathbb X)$ denotes the maximum norm of elements in $\mathbb X$, namely $\varrho(\mathbb X):=\sup_{x\in \mathbb X}\|x\|$.
\subsection{Working Assumptions} 
\begin{assumption}\label{assf0positive}
The cost function value $f_0(p)$ is nonnegative for all $p$.\ \\ 
\end{assumption}
This assumption can be made satisfied by adding sufficiently high positive constant. It is quite common in MPC context where the cost function refers quite often to the integral of the tracking error that is added to some positive terminal term. \ \\ 
\begin{assumption}\label{asslyp} 
There are two reals $L_0\ge 0$ and $L_\psi\ge 0$ such that 
$f_0\in \mathcal F^1_{L_0}$ and $\psi\in \mathcal F^1_{L_\psi}$ \ \\ 
\end{assumption}
\begin{assumption}\label{assconv}
There is $\mu_0>0$ such that $f_0\in \mathcal S^1_{\mu_0}$. Moreover, $\Psi$ is convex.
\end{assumption}
\ \\
Note that this assumption implies that $f\in \mathcal S^1_{\mu_0}$ and that there is a unique critical point for $f$ which is denoted hereafter by $p^*\in \mathcal C$. therefore according to the definition of $d(p)$, one has $d(p):=\|p-p^*\|$. \ \\ \ \\  
In what follows, the notation $p_u$ and $p_a$ refer to two vectors such that:
\begin{eqnarray}
p_u:=\min_{p\in \mathbb{R}^{p}}f_0(p)\quad;\quad \psi(p_a)\le 0
\end{eqnarray} 
namely, $p_u$ is the unconstrained minimum of $f_0$ while $p_a$ is any admissible point. Having $p_a$, the following definition can be stated since $f_0$ is supposed to be continuously differentiable and because $f_0\in \mathcal S^1_{\mu_0}$ [Assumption \ref{assconv}]:\ \\
\begin{definition} \label{defdeD0Dpsi} 
Define $D_0$ by:
\begin{eqnarray}
D_0:=\sup_{f_0(p)\le f_0(p_a)} \| f_0^{'}(p)\| \ge 0
\end{eqnarray} 
\end{definition}  
\begin{remark} \label{rempa} 
In fact, the knowledge of the admissible point $p_a$ is only required to compute $D_0$. therefore, if an upper bound of $D_0$ can be found, the knowledge of $p_a$ is not mandatory. This is clearly shown in section \ref{tgFFFF65} in the specific case of QP problems [see inequality (\ref{rftdrefRR})] . This is crucial since in the MPC context the constraints are state dependent and it may become cumbersome to compute $p_a$ for each current state.   
\end{remark}
\ \\
The next assumption concerns the behavior of the penalty map outside the admissible set.\\  
\begin{assumption} \label{assphi} 
There is $\beta>0$ such that the following inequality:
\begin{eqnarray}
\psi(p)\ge \beta\times \Bigl[d(p,\mathcal A_{\psi=0})\Bigr]^2 \label{defdebeta} 
\end{eqnarray} 
holds for all $p$. $\hfill \diamondsuit$
\end{assumption}
The expressions of the parameters $L_0$, $L_\psi$, $\mu_0$, $D_0$ and $\beta$ in the specific case of quadratic cost $f_0$ and affine in $p$ constraints $c_i$ are given in section \ref{tgFFFF65}. 
\subsection{Preliminary results}
\noindent In this section some preliminary results are stated. For better readability, all the proofs are given in the appendix. 
The first result gives a property of the gradient of $f_0$ at the stationary point $p^*$:\\ 
\begin{lemma} \label{lemmapropderive} 
The following inequality holds
\begin{eqnarray}
\|f_0^{'}(p^*)\|\le D_0
\end{eqnarray} 
\end{lemma}
{\sc Proof}. See Appendix \ref{prooflemmapropderive}. \ \\ \ \\  
The following result characterizes the behavior of the penalty term $\psi$ in terms of the penalty coefficient $\rho$:\\ 
\begin{lemma}\label{propexpressphi} 
{\bf If} $\rho>L_0/\beta$ 
{\bf then}  the following inequality:
\begin{eqnarray}
\psi(p)\le \dfrac{L_\psi}{2}\Bigl[d(p)+\dfrac{\kappa_0}{\sqrt{\rho}}\Bigr]^2 \  \mbox{\rm where}\ \kappa_0:=\dfrac{2L_0}{\beta}\sqrt{\dfrac{2}{\mu_0}\psi(p_u)}\nonumber \ \\ \label{hgYYg6} 
\end{eqnarray} 
holds for all $p$. In particular, for $p^*$ one has:
\begin{eqnarray}
\psi(p^*)\le \dfrac{L_\psi \kappa_0^2}{2\rho} \label{oju76756} 
\end{eqnarray} 
\end{lemma} 
{\sc Proof} See Appendix \ref{lemmapropexpressphi}. \ \\ \ \\   
Note that Lemma \ref{propexpressphi} quantifies how increasing $\rho$ leads to a smaller constraint violation depending on the amount of violation $\psi(p_u)$ at the unconstrained minimum $p_u$ of $f_0$. \ \\ \ \\ 
The following corollary gives a bound on the difference in the cost $f_0$ evaluated at the unconstrained optimum $p^*$ of $f=f_0+\rho\psi$ and the true optimal cost as a function of constraint violation: \\
\begin{lemma} \label{corcorcorhg6} 
Let $p^{opt}$ be the optimal solution of the original problem (\ref{defproblem}). $p^*$ the unconstrained minimum of $f$. The following inequality holds:
\begin{eqnarray}
\vert f_0(p^{opt})-f_0(p^*)\vert \le D_0\left[\dfrac{\psi(p^*)}{\beta}\right]^{\frac{1}{2}}+\dfrac{L_0}{2}\left[\dfrac{\psi(p^*)}{\beta}\right] \label{jhGG034} 
\end{eqnarray} 
\end{lemma}
{\sc Proof}. See Appendix \ref{proofcorcorcorhg6}. \ \\ \ \\   
Using Lemma \ref{corcorcorhg6} one can prove the following result:\\
\begin{corollary}
{\bf If} the following inequality holds: 
\begin{eqnarray}
\left[\frac{\psi(p^*)}{\beta}\right]^{\frac{1}{2}}\le Z_1(\epsilon):=\dfrac{D_0}{L_0}\left[\left(1+\frac{2L_0}{D_0^2}\epsilon\right)^{\frac{1}{2}}-1\right] \label{defdeZZZ1} 
\end{eqnarray} 
{\bf then} the stationary solution $p^*$ satisfies:
\begin{eqnarray}
\vert f_0(p^{opt})-f_0(p^*)\vert \le \epsilon
\end{eqnarray}  
\end{corollary}
{\sc Proof}.  
This can be easily obtained after noticing that the r.h.s of (\ref{jhGG034}) is a second order polynomial in $\sqrt{\psi(p^*)/\beta}$. Writing that this polynomial is equal to $\epsilon$ and solving for it gives the result. $\hfill \Box$\ \\ \ \\ 
Note however that $p^*$ is never reached exactly. Instead, the fast gradient iteration will be used to reach an iterate $p$ that is close to $p^*$. Now since the available certification bounds on the fast gradient iterations concern the guaranteed value of $\vert f(p^*)-f(p)\vert$ while the $\bar\varepsilon$-suboptimality is defined in terms of the original cost $f_0$, the following lemma gives a link between these two indicators:\\
\begin{lemma} \label{fgfYH} 
The following implication holds for all $\epsilon$:
\begin{eqnarray}
\Bigl\{\vert f(p)-f(p^*)\vert\le \epsilon\Bigr\} \Rightarrow \nonumber \\ 
\Bigl\{\vert f_0(p)-f_0(p^*)\vert\le D_0\left[\frac{2\epsilon}{\mu_0}\right]^{\frac{1}{2}}+\dfrac{L_0}{2}\left[\frac{2\epsilon}{\mu_0}\right]\Bigr\} \label{onze} 
\end{eqnarray} 
\end{lemma}
{\sc Proof}. See Appendix \ref{prooffgfYH}. \ \\ \ \\   
Here again, Lemma \ref{fgfYH} gives the condition on the precision $\epsilon_1$ required on $f$ in order to induce a precision $\epsilon_2$ on $f_0$, namely:\\
\begin{corollary}
{\bf If} $p$ is such that $\vert f(p)-f(p^*)\vert \le \epsilon_1$  with
\begin{eqnarray}
\left[\dfrac{2\epsilon_1}{\mu_0}\right]^{\frac{1}{2}}\le Z_1(\epsilon_2)
\end{eqnarray} 
where $Z_1$ is the function defined by (\ref{defdeZZZ1}) {\bf then}, one has $\vert f_0(p)-f_0(p^*)\vert \le \epsilon_2$. 
\end{corollary}
\ \\
{\sc Proof}. Use the same arguments as before since the r.h.s of (\ref{onze}) involves the same polynomial as in (\ref{jhGG034}). $\hfill \Box$\ \\ \ \\ 
The certification bound of the fast gradient needs an upper bound on the distance between the initial guess $p$ and the minimizer of $f$, namely $p^*$. The following lemma gives such an upper bound in terms of the value of the function $f$ at the initial guess $p$:\\ 
\begin{lemma} \label{lemPstarrho}
The following inequality is satisfied for all $p$:
\begin{eqnarray}
\|p^*-p\|\le \left[\dfrac{2f(p)}{\mu_0}\right]^{\frac{1}{2}}=:r(p) \label{defderrrr} 
\end{eqnarray} 
\end{lemma}
{\sc Proof}. This is a direct consequence of the inclusion $f\in \mathcal S_{\mu_0}^1$ and the fact that $f_0$ (and hence $f$) is positive. $\hfill \Box$
\section{The Algorithm} \label{secalgo} 
\subsection{Recalls on the Fast Gradient iteration}
\noindent The fast gradient algorithm proposed in \cite{Nesterov2004} is commonly  used to perform unconstrained minimization of a function  $f\in \mathcal S_{\mu,L}$. It is briefly recalled through Algorithm \ref{figFG} for which the following convergence result holds
\begin{algorithm}
\caption{$[p_N,q_N,\alpha_N]=F^{(N)}(p_0,q_0,\alpha_0)$}\label{figFG} 
\begin{algorithmic}[1]
    \For{$i=1:N$} 
    \State$p_{i+1}\leftarrow q_i-f^{'}(q_i)/L$
    \State Compute $\alpha_{i+1}\in (0,1)$ solution of $\alpha_{i+1}^2=(1-\alpha_{i+1})\alpha_i^2+\mu_0\alpha_{i+1}/L$
    \State $\beta_i\leftarrow \left(\alpha_i(1-\alpha_i)\right)/(\alpha_i^2+\alpha_{i+1})$
    \State $q_{i+1}\leftarrow p_{i+1}+\beta_i(p_{i+1}-p_i)$
    \EndFor
\end{algorithmic}
\end{algorithm}
\begin{propos}(\cite{Nesterov2004}, page 80) \label{propos1}
The successive iterates of Algorithm \ref{figFG} starting from the initial guess $p_0$, $\alpha_0=\sqrt{\mu_0/L}$ and $q_0=p_0$ satisfy the following inequality:
\begin{eqnarray}
f(p_i)-f(p^*)\le \nonumber \\ \dfrac{L+\mu_0}{2}\times \min\left\{(1-c)^i,\dfrac{1}{(1+ic)^2}\right\} \times \|p_0-p^*\|^2 \label{uj17} 
\end{eqnarray} 
where $c:=\sqrt{\mu_0/L}$ and where $p^*$ stands for the unconstrained minium of $f$. $\hfill \diamondsuit$
\end{propos}
\ \\ 
The following is a direct consequence of Proposition \ref{propos1}:\\
\begin{corollary}\label{corcor9}
If the initial guess satisfies $\|p_0-p^*\|\le \delta$ then for any $\epsilon>0$, the integer:
\begin{eqnarray}
\bar N(c,\gamma):=\max\left\{0,\min\{\dfrac{\log(\gamma)}{\log(1-c)},\dfrac{1}{c}(\sqrt{\dfrac{1}{\gamma}}-1)\}\right\}\label{defdeNbarbar}\\
\mbox{\rm where}\quad  \gamma:=2\epsilon/((L+\mu_0)\delta^2)\ ;\  c=\sqrt{\mu_0/L}  \nonumber 
\end{eqnarray}  
is an upper bound of the number of iterations $N$ needed by Algorithm \ref{figFG} to deliver a sub-optimal solution $p_N$ satisfying $\vert f(p_N)-f(p^*)\vert\le \epsilon$. $\hfill \diamondsuit$
\end{corollary}
\ \\
{\sc Proof}. Inject $\|p_0-p^*\|\le \delta$ in (\ref{uj17}) and impose that the r.h.s is $\le\epsilon$. $\hfill \Box$\ \\ 
Now using the bound on $\|p_0-p^*\|\le r(p_0)$ given by (\ref{defderrrr}), the following result follows:\\
\begin{corollary}\label{cortg6} 
Given any initial value $p_0$, let $\gamma_0:=\epsilon\mu_0/[(L+\mu_0)f(p_0)]$ then
$N(c,\gamma_0)$ is an upper bound of the number of iterations $N$ needed by Algorithm \ref{figFG} to deliver a sub-optimal solution $p_N$ satisfying $\vert f(p_N)-f(p^*)\vert\le \epsilon$. $\hfill \diamondsuit$
\end{corollary}
\subsection{The  Proposed Algorithm}
\noindent The proposed algorithm involves the quantities defined by (\ref{defdesrhos})-(\ref{defdeseta}) that depend on: 
\begin{itemize}
\item the problem's intrinsic properties $(\mu_0,L_0,L_\psi,\beta,D_0)$
\item the unconstrained solution-dependent parameter $\kappa_0$ [see (\ref{hgYYg6})] 
\item the desired precision pair $\bar\varepsilon:=(\varepsilon_0,\varepsilon_\psi)$
\end{itemize} 
\begin{minipage}{0.22\textwidth}
\small
\begin{align}
\begin{split}
\rho_1&:=\dfrac{2L_\psi\kappa_0^2}{\varepsilon_\psi^2}\\
\rho_2&:=\dfrac{L_\psi\kappa_0^2}{2\beta Z_1^2(\varepsilon_0/2)}\\
&\ \\
\rho_3&:=L_0/\beta
\end{split}\label{defdesrhos} 
\end{align}  
\end{minipage} 
\begin{minipage}{0.26\textwidth}
\small
\begin{align}
\begin{split}
\eta_1&:=\dfrac{\mu_0}{2}Z_1^2(\dfrac{\varepsilon_0}{2})\\
\eta_2&:=\frac{\mu_0\varepsilon_\psi^2}{4L_\psi}
\end{split}\label{defdeseta}
\end{align}
\end{minipage} 
\ \\ \ \\ 
These  quantities are used in Algorithm \ref{figFGproposed} below:
\begin{algorithm} \label{A2A2A2} 
\caption{$\hat p^*=A(p_0,\bar\varepsilon:=(\varepsilon_0,\varepsilon_\psi))$}\label{figFGproposed} 
\begin{algorithmic}[1]
    \State $\alpha_0=(\mu_0/L)^{\frac{1}{2}}$, $q_0:=p_0$
    \State $\rho=\max\{\rho_1,\rho_2,\rho_3\}$
    \State $\eta=\min\{\eta_1,\eta_2\}$
    \State $c=\sqrt{\mu_0/L}$
    \State $\gamma_0=\eta\mu_0/[(L+\mu_0)f_0(p_0)]$
    \State $N_{max}=\bar N(c,\gamma_0)$
    \State $g_{min}=\mu_0\sqrt{2\eta/L}$
    \State again=true
    \While{(again)} 
    \State $[p_{i+1},q_{i+1},\alpha_{i+1}]=FG^{(1)}(p_{i},q_{i},\alpha_{i})$
    \If{[$(i\ge N_{max})$ or $(g(p_i)\le g_{min})$]}
    \State again=0
    \Else
    \State $i=i+1$
    \EndIf
    \EndWhile
    \State $\hat p^*=p_{i}$
\end{algorithmic}
\end{algorithm}
\ \\
The following result gives a certification bound on the number of iterations needed by Algorithm \ref{figFGproposed} to achieve an $\bar\varepsilon$-suboptimal solution of the original problem.  \\
\begin{propos} \label{mainresult} 
Let be given a precision pair $\bar\varepsilon:=(\varepsilon_0,\varepsilon_\psi)$, an initial guess $p_0$. Let $\gamma_0:=\eta\mu_0/[(L+\mu_0)f_0(p_0)]$ where $\eta:=\min\{\eta_1,\eta_2\}$ with the $\eta_i$s given by (\ref{defdeseta}). The algorithm in which $\rho=\max\{\rho_1,\rho_2,\rho_3\}$ is used with the $\rho_i$s defined by (\ref{defdesrhos})  involves at most $\bar N(c,\gamma_0)$ unconstrained fast gradient elementary iterations before it delivers an estimate $\hat p^*$ that is an $\bar\varepsilon$-suboptimal solution of the original constrained optimization problem (\ref{defproblem}).  
\end{propos}
\ \\
{\sc Proof}. See Appendix \ref{proofmainresult}. \ \\ \ \\ 
In the remainder of the paper, the maximum number of iterations that guarantee the precision as expressed in Proposition \ref{mainresult} is denoted by:
\begin{eqnarray}
N(p_0,\varepsilon_0,\varepsilon_\psi):=\bar N(c,\gamma_0) \label{defdenN} 
\end{eqnarray} 
as the arguments of $N$ completely determine $c$ and $\gamma_0$.
\subsection{Case of Quadratic Programming (QP) problems} \label{tgFFFF65} 
\begin{figure}
\begin{center}
\input{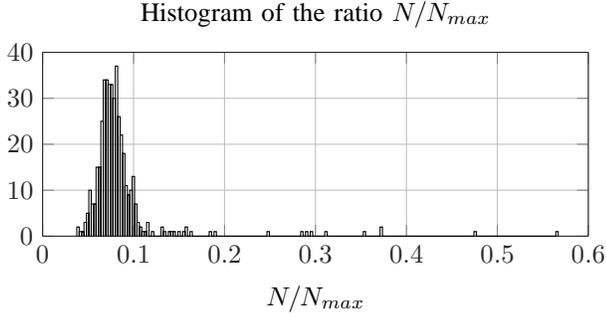}
\end{center} 
\caption{Histogram showing the statistics of the ratio $N/N_{max}$ between the effectively needed number of iterations $N$ and the certification bound $N_{max}$ computed from the theory when using the numerical experiments described in section \ref{secvalidQPseul}.} \label{fighisto} 
\end{figure}

\noindent Here, the expressions of $L_0$, $L_\psi$, $\mu_0$, $D_0$ and $\beta$ are given in the specific case of QP problems where the cost function and the constraints take the form:
\begin{eqnarray*}
f_0(p)=\dfrac{1}{2}p^THp+F^Tp+s_0\quad ;\quad c_i(p)=A_ip-B_i
\end{eqnarray*} 
In this case, Assuming that $s_0$ is such that assumption \ref{assf0positive} holds, it is straightforward that  Assumptions \ref{asslyp} and \ref{assconv} holds with $L_0=\lambda_{max}(H)$, $L_\psi=\sigma_{max}(A)$ and $\mu_0=\lambda_{min}(H)$. Moreover, one has $p_u:=-H^{-1}F$. Now according to remark \ref{rempa}, $p_a$ is not needed provided that an upper bound for $D_0$ can be derived. This is the aim of the following proposition:
\begin{propos} \label{propD0} 
Provided that the set of inequalities $Ap\le B$ implies the condition $p\in \mathbb P$, the following inequality holds:
\begin{eqnarray}
D_0\le &&\left[\lambda_{max}(H)\right]\cdot \bar p+\|F\| \label{rftdrefRR} 
\end{eqnarray} 
where 
\begin{eqnarray}
\bar p&:=&\dfrac{\|F\|+\sqrt{\|F\|^2+2\lambda_{min}(H)\left[\bar f\right]}}{\lambda_{min}(H)} \label{defdepbar} 
\end{eqnarray} 
in which 
\begin{eqnarray}
\bar f:=\dfrac{1}{2}\lambda_{max}(H)\left[\varrho(\mathbb P)\right]^2+\|F\|\cdot \varrho(\mathbb P)\label{defdefbar} 
\end{eqnarray} 
\end{propos}
{\sc Proof}. See Appendix \ref{proofpropD0}. \ \\ \ \\ 
Assumption \ref{assphi} is satisfied with $\beta:=\sigma_{min}(A)$ which is the lowest non zero singular value of the constraints matrix $A$. The coefficient $\kappa_0$ involved in lemma \ref{propexpressphi} and the expressions (\ref{defdesrhos})-(\ref{defdeseta}) used to compute $\rho$ and $\eta$ is obtained using the values of $L_0$, $\beta$, $\mu_0$ and $p_u$ mentioned above.  \ \\ \ \\ 
\underline{\sc Numerical experiments} \label{secvalidQPseul} 
\noindent In order to check the validity of the certification bound $N(p_0,\varepsilon_0,\varepsilon_\psi)$, $500$ random QP problems have been generated with $n=10$ decision variables and $n_c=20$ constraints. More precisely,  $H:=CC^T+\sigma \mathbb I$ is used where $C\in \mathbb{R}^{n\times 1}$ and $\sigma\in [10^{-3},1]$, $F$ and $s_0$ has been computed so that the cost is $\|p-p_{u}\|_H^2+1$ where $p_u$ is randomly generated. The constraints matrices $A\in \mathbb{R}^{n\times n_c}$ and $B\in \mathbb{R}^{n_c}$ has been randomly generated so that a feasible solution exists. The precision $\epsilon_\psi=10^{-2}$ has been used while $\varepsilon_0$ has been systematically taken equal to $1\%$ of the true optimal cost that is obtained by {\sc quadprog-Matlab} solver. The initial guess is systematically taken equal to $0$ as one might use in cold start MPC context. \ \\ \ \\ The results are shown in Figure \ref{fighisto} where the histogram over the $500$ trials of the ratio between the effectively needed number of iterations $N$ and the maximal computed certification bound $N_{max}$ (step 6 of Algorithm 2) is plotted. The results suggest that for this class of QP problems, the bounds is not that conservative and that since some scenarios lead to a ratio between 0.5 and 0.6, as far as certification is needed, it cannot be strongly reduced. 

\section{Application to Real-Time MPC} \label{apprtMPC} 
\noindent In this section, it is assumed that a certification bound $N(p_0,\varepsilon_0,\varepsilon_\psi)$ is given for some algorithm. Based on such a bound, a real-time MPC implementation framework is proposed using a state-dependent control updating period leading to provable practical convergence. It is therefore important to underline that the results of this section does not necessarily relate to the use of the fast-gradient algorithm as they can apply to any algorithm for which a certification can be associated that depends on the initial guess $p_0$ and some required precision pair $(\varepsilon_0$ and $\varepsilon_\psi$) in the sense of (\ref{bbHGFF}). 
\subsection{Definition, notation and working assumptions}
\noindent In this section, a set of assumptions are stated. Not all of them are used in all the subsequent results. That is why in the statement of each result, the assumptions that are needed are explicitly mentioned. \ \\ \ \\ 
In MPC framework, the controller disposes of a model of the form
\begin{eqnarray}
\dot x=F(x,u) \quad (x,u)\in \mathbb{R}^{n}\times \mathbb{R}^{n_u}\label{xdotdot} 
\end{eqnarray} 
where the following assumption is used regarding the definition of the vector $x$:\\
\begin{assumption} \label{asswx} 
The state vector $x$ involved in (\ref{xdotdot}) gathers the physical state of the system together with the current set-point and current estimation of the disturbance. The model also incorporates the assumption on the future behavior of these exogenous variables.
\end{assumption}
\ \\
We consider that the future control profiles are parametrized through a finite dimensional vector $p$ of degrees of freedom such that at each instant $t$, the future profile depends on $p(t)$ according to:
\begin{eqnarray}
u(t+s):=\mathcal U(s,p(t))\quad s\in [0,T] \label{defdeuparam} 
\end{eqnarray} 
where $\mathcal U$ is some predefined map and $T$ is the prediction horizon. \ \\ \ \\ 
Since the MPC has to be computed based on the prediction of the future state (in the sense of Assumption \ref{asswx}), the following assumption is needed to characterize the state prediction error:\\ 
\begin{assumption}\label{assprederror} 
For each compact set $\mathbb C$ to which belongs the pair $(p(t),x(t))$, the prediction $\hat x(t+\tau)$ of the future state starting from $x(t)$ and under the control profile $\mathcal U(\cdot,p(t))$ can be affected by an error satisfying\begin{eqnarray}
\|\hat x(t+\tau)-x(t+\tau)\|\le E^0_\mathbb{C}+E^1_\mathbb{C}\times \tau \label{defdeE0E1} 
\end{eqnarray} 
\end{assumption}
\ \\
Note that $E^0_\mathbb{C}$ in (\ref{defdeE0E1}) accommodates for unpredictable set-point changes while $E^1_\mathbb{C}$ accommodates for the presence of disturbances that affects the input of some integrator in the system or for the presence of unpredictable  time-varying set-point. \ \\ \ \\ 
The cost function is defined at instant $t$ based on the knowledge of the state $x(t)$ (including the current value of the set point and the disturbance estimation and prediction). This leads to a constrained optimization problem of the form (\ref{defproblem})  in which both $f_0$ and $c_i$ are dependent on the current value $x(t)$ of the state, namely:
\begin{eqnarray*}
f_0(p,x(t))\quad;\quad \psi(p,x(t))
\end{eqnarray*}  
Consequently, the call of Algorithm \ref{figFGproposed}  as well as the bound (\ref{defdenN})  on the number of iterations must now incorporate the state $x(t)$ as an argument, namely:
\begin{eqnarray}
\hat p^*=A(p_0,\varepsilon_0,\varepsilon_\psi,x)\quad;\quad  N(p_0,\varepsilon_0,\varepsilon_\psi,x) \label{hghghg2365} 
\end{eqnarray} 
In order to use the results of the preceding section, one needs to assume that for all $x$, there are positive reals $L_0(x)$, $L_\psi(x)$ and $\beta(x)$ and a strictly positive $\mu_0(x)>0$ that play the roles of $L_0$, $L_\psi$, $\beta$ and $\mu_0$ as defined in the preceding section. \ \\ \ \\ 
Now if for some reasons, one knows that the pair $(p_0,x)$ involved in (\ref{hghghg2365})  belongs to some compact set $\mathbb C:=\mathbb P\times \mathbb X$, then one can obtain a certification bound that depends only on the precision parameters $\bar\varepsilon:=(\varepsilon_0,\varepsilon_\psi)$, namely:
\begin{eqnarray}
N_\mathbb{C}(\varepsilon_0,\varepsilon_\psi):=\max_{(p,x)\in \mathbb C} N(p,\varepsilon_0,\varepsilon_\psi,x) \label{nghghf} 
\end{eqnarray}   
Moreover, the following result shows that the bound $N_\mathbb{C}(\varepsilon_0,\varepsilon_\psi)$ can be computed through static optimization steps involving the functions $f_0$ and $\psi$:
\begin{propos} \label{ygtf7865} 
Let a compact set $\mathbb C:=\mathbb P\times \mathbb X$ be given. the bound $N_\mathbb{C}(\epsilon_0,\epsilon_\psi)$ defined by (\ref{nghghf}) can be computed by the following steps:
\begin{enumerate}
\item Compute $\psi^{max}$ according to:
\begin{eqnarray}
\psi^{max}:=\max_{x\in \mathbb X}\Bigl\{\psi(p_u,x)\ \vert\ f_0^{'}(p_u,x)=0\Bigr\} \label{defdepsimaxhaut} 
\end{eqnarray} 
\item Compute $L_0$, $L_\psi$ as the maximum of $L_0(x)$ and $L_\psi(x)$ over $x\in \mathbb X$
\item Compute $\beta$ and $\mu_0$ as the minimums of $\beta(x)$ and $\mu_0(x)$ over $x\in \mathbb X$
\item Compute $\kappa_0^{max}:=\dfrac{2L_0}{\beta}\sqrt{2\psi^{max}/\mu_0}$
\item Compute $\rho^{max}$ using (\ref{defdesrhos}) in which $\kappa_0^{max}$ replaces $\kappa_0$ 
\item Compute $\eta^{min}:=\min\{\eta_1,\eta_2\}$ where the $\eta_i$ are computed by (\ref{defdeseta}) in which $\rho^{max}$ replaces $\rho$. 
\item Compute $f_0^{max}:=\max_{(p,x)\in \mathbb C}f_0(p,x)$
\item Compute $\gamma_0^{min}:=\eta^{min}\mu_0/[(L(\rho^{max})+\mu_0)f_0^{max}]$
\item Compute $c^{min}:=\sqrt{\mu_0/L(\rho^{max})}$
\end{enumerate}
Finally compute the desired quantity:
\begin{eqnarray}
N_\mathbb{C}(\varepsilon_0,\varepsilon_\psi):=\bar N(c^{min},\gamma_0^{min})
\end{eqnarray}  
where $\bar N$ is defined by (\ref{defdeNbarbar}).
\end{propos}
{\sc Proof}. Straightforward as the computation systematically takes the worst case towards the increase of $N$. $\hfill \Box$ \ \\ \ \\ 
In section \ref{QPMPCCCC}, Explicit computation of all the quantities involved in Proposition \ref{ygtf7865} is given for the specific case of state-dependent QP optimization problems that arise in the linear MPC context. \ \\ \ \\  
It is also assumed that the cost function $f_0$ is proper in both $p$ and $x$ in the following sense:\\
\begin{assumption} \label{assCphi} 
For any positive real $\phi>0$, there is a compact set $\mathbb C_\phi$ such that the following implication holds:
\begin{eqnarray}
\Bigl\{f_0(p,x)\le \phi\Bigr\}\Rightarrow \Bigl\{(p,x)\in \mathbb C_\phi\Bigr\} \label{defdeCphibarbarbar} 
\end{eqnarray} 
\end{assumption}

\begin{figure}

\begin{center}
\begin{tikzpicture}[scale=0.7]
\draw[<->,>=stealth] (5,0) node[below]{\footnotesize $s$} -- (0,0) -- (0,3) node[right, RubineRed]{\footnotesize $\ell(s,p,x)$};
\node[left] at (0,2){\footnotesize $q(x)$};
\draw[fill] (0,2) circle (1pt);
\node[below] at(1.5,0) (O1){\footnotesize $\dfrac{q(x)}{D_{\mathbb C}}$};
\draw[very thick,Blue] (0,2) -- (1.5,0) -- (4.8,0);
\draw[RubineRed,thick] plot[smooth] coordinates {(0,2.2) (1,1) (2,0.2) (3,0.1) (5,0)};
\end{tikzpicture}
\end{center} 
\caption{Illustration of Assumption \ref{asslC}.} \label{figillusell} 
\end{figure}
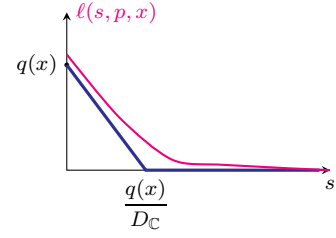

Regarding the dependence of $f_0$ and $\psi$ on $x$, the following assumption is considered:\\ 
\begin{assumption} \label{assdx} 
For any compact set $\mathbb C$, there are positive real $K_\mathbb{C}^0, K_\mathbb{C}^\psi>0$ such that :
\begin{eqnarray}
\|f_0(p,x_1)-f_0(p,x_2)\|\le K_\mathbb{C}^0\cdot \|x_1-x_2\| \label{defdeKdep} \\
\|\psi(p,x_1)-\psi(p,x_2)\|\le K_\mathbb{C}^\psi\cdot \|x_1-x_2\| \label{defdeKdepsi} 
\end{eqnarray} 
for all $(p,x_1), (p,x_2)\in \mathbb C$.
\end{assumption}
\ \\
A typical formulation of $f_0(p,x_0)$ in MPC is given by:
\begin{eqnarray}
f_0(p,x_0)&:=&\Omega(\bar x(T,p,x_0))+\int_0^T \ell(\bar x(s,p,x_0),p,s)ds\nonumber \\
&=:&\Omega(\bar x(T,p,x_0))+\int_0^T\bar \ell(s,p,x_0)ds \label{de548OI} 
\end{eqnarray} 
where $\bar x(s,p,x_0)$ is the predicted state value at instant $s$ starting from $x_0$ at instant $0$. \ \\ \ \\ Regarding the formulation of the MPC, the following (commonly satisfied) assumption is needed in the sequel:
\begin{assumption} \label{assDelta} 
The MPC formulation is based on a cost function of the form (\ref{de548OI}) with the necessary constraints that make the following inequality satisfied:
\begin{eqnarray}
&&f_0(p^{opt}(t+\tau),\hat x(t+\tau))-f_0(p^{opt}(t),x(t))\le \nonumber \\ &&\le -\Delta(\tau,x(t)):=-\int_0^\tau \bar \ell(s,p^{opt}(t),x(t))ds \label{gfTTRTRDE} 
\end{eqnarray} 
where $p^{opt}(t)$ is the optimal solution of the problem defined for the state $x(t)$ while $p^{opt}(t+\tau)$ is the optimal solution of the problem defined by the predicted future state $\hat x(t+\tau)$ starting from $x(t)$ under the optimal control $\mathcal U(\cdot,p^{opt}(t))$ that is applied on the interval $[t,t+\tau]$.  
\end{assumption}
\ \\
Note that $p^{opt}(t)$ does not appear as an argument of $\Delta$ since $p^{opt}(t)$ is assumed to be uniquely determined by $x(t)$.
\begin{remark}
Note that the inequality (\ref{gfTTRTRDE}) is satisfied only for the ideal predicted future state $\hat x(t+\tau)$ since otherwise the bad knowledge of uncertainties and/or the set-point changes may invalidate the inequality if the true value $x(t+\tau)$ of the state is used. 
\end{remark}
\begin{remark}
Note that inequality (\ref{gfTTRTRDE}) is commonly satisfied in the standard provably stable MPC formulations. Moreover, the r.h.s $\Delta(\tau,x(t))$ is generally exhibited through the corresponding stability proof (see \cite{Mayne:2000}). 
\end{remark}
Regarding the penalty function $\ell$, the following assumption is used:\\
\begin{assumption} \label{asslC} 
[Figure \ref{figillusell}] For any compact set $\mathbb C$, there is a positive real $D_\mathbb{C}>0$ and a positive function $q(\cdot)$ such that :
\begin{eqnarray}
\bar \ell(s,p,x)\ge \max\left\{0,q(x)-D_{\mathbb C}s\right\}\label{qx} 
\end{eqnarray} 
for all $(p,x)\in \mathbb C$.
\end{assumption} 
\ \\ 
Note that condition (\ref{qx}) simply states that with bounded control, there is a limitation on the speed with which the state can be steered to the desired region. With this respect, $q(x)$ is simply a state dependent term in $\ell$ that expresses how far does $x$ lie from the desired region. This notation enables many situations to be handled as $x$ includes set-point definition and therefore, mesures of the difference between the physical state of the system and their desired value can take the simple form expressed by $q(x)$.\ \\ \ \\ 
Finally, the following assumption is used to characterize the available computational facility:\\
\begin{assumption}\label{asstauc} 
The system is controlled with a computational facility that performs a single elementary iteration of the fast gradient (step 9 of Algorithm \ref{figFGproposed}) in $\tau_c$ time units. 
\end{assumption}
\ \\
Note that if another certified algorithm than the fast gradient is used, $\tau_c$ used hereafter denotes the time necessary to perform a single iteration of that specific algorithm. 
\subsection{Certified MPC by state-dependent updating period}
Assume that a scheme is based on the iterative on-line definition of a sequence of updating instants and a sequence of precision parameters denoted by:
\begin{eqnarray}
t_{k+1}=t_k+\tau_k\quad;\quad \bigl\{\varepsilon_0^{(k)},\varepsilon_\psi^{(k)}\bigr\}_{k=0}^\infty
\end{eqnarray} 
which are linked through the definition of the updating periods $\tau_k$ according to:
\begin{eqnarray}
\tau_k:=\tau_c\times N_\mathbb{C}\bigl(\varepsilon_0^{(k+1)},\varepsilon_\psi^{(k+1)}\bigr) \label{defdetauk} 
\end{eqnarray} 
where $\mathbb{C}$ is some compact subset of $\mathbb R^{n_p}\times \mathbb R^n$ and $\tau_c$ is the computation time needed for a single fast gradient iteration (see Assumption \ref{asstauc}). \ \\ \ \\ More precisely, given the current state $x(t_k)$ and a control $\mathcal U(\cdot,\hat p^*(t_k))$ that is applied during the sampling period $[t_k,t_{k+1}]$, Algorithm \ref{figFGproposed} is used to compute the control parameter $\hat p^*(t_{k+1})$ (that is to be applied during the next sampling period) with the hot start $[\hat p^*(t_k)]^{+\tau_k}$ and the precision parameters $(\varepsilon_0^{(k+1)},\varepsilon_\psi^{(k+1)})$. Note that by the very definition (\ref{defdetauk}) of $\tau_k$, the value of the control parameter $\hat p^*(t_{k+1})$ that is obtained by Algorithm \ref{figFGproposed} before $t_{k+1}$ necessarily meets the precision requirements, namely:\begin{eqnarray}
&&f_0(\hat p^*(t_{k+1}),\hat x(t_{k+1}))-f_0(p^{opt}(t_{k+1}),\hat x(t_{k+1}))\le \varepsilon_0^{(k+1)}\nonumber \\ 
&&c_i(\hat p^*(t_{k+1}),\hat x(t_{k+1}))\le 0\qquad \quad i\in I_h \label{ingt77}\\
&&c_i(\hat p^*(t_{k+1}),\hat x(t_{k+1}))\le \varepsilon_\psi^{(k+1)}\quad i\in I_s \nonumber 
\end{eqnarray} 
Using the first inequality, one can prove the following result:\\
\begin{lemma} \label{lemfondineq} 
{\bf If} the following conditions hold 
\begin{enumerate}
\item $\tau_k$ is defined by (\ref{defdetauk}) for some compact set $\mathbb C:=\mathbb P\times \mathbb X$
\item For all $k$, $[\hat p^*(t_k)]^{+\tau_k}\in \mathbb P$
\item For all $k$, $x(t_k)\in \mathbb X$
\item Assumptions \ref{assprederror}, \ref{assdx}  and \ref{assDelta} are satisfied 
\end{enumerate}  
{\bf then} the following inequality holds for all $k$:
\begin{eqnarray}
&&f_0(\hat p^*(t_{k+1}),x(t_{k+1}))-f_0(\hat p^*(t_{k}),x(t_{k}))\le \nonumber \\
&&\varepsilon_0^{(k)}+K^0_\mathbb{C}(E^0_\mathbb{C}+E^1_\mathbb{C}\tau_k)+\varepsilon_0^{(k+1)}-\Delta(\tau_k,x(t_k)) \label{lhs45} 
\end{eqnarray}  
\end{lemma}

\begin{figure}
\begin{center}
\begin{tikzpicture}
\draw[<->,>=stealth] (5,0) node[below]{\footnotesize $\tau$} -- (0,0) -- (0,2) node[right]{\footnotesize $\Gamma_\mathbb{C}(\tau,q)$};
\node[below] at(2,0) (O1){\footnotesize $q/D_{\mathbb C}$};
\draw[fill] (2,0) circle (1pt);
\node[left] at(0,1) (O2){\footnotesize $q^2/(2D_{\mathbb C})$};
\draw[fill] (0,1) circle (1pt);
\draw[dotted,thin] (O1) |- (O2);
\draw[scale=1,domain=0:2,smooth,variable=\x,blue,thick] plot ({\x},{\x-0.25*\x*\x}) -- (4.8,1);
\end{tikzpicture}
\end{center} 
\caption{Evolution of $\Gamma_\mathbb{C}(\tau,q)$ involved in Lemma \ref{simplergloballemma}.}\label{figGamma} 
\end{figure}
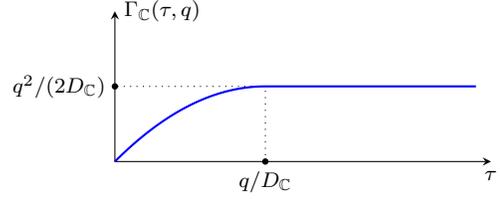

{\sc Proof}. See Appendix \ref{prooflemfondineq}.  \ \\ \ \\ 
Note that the term $f_0(\hat p^*(t_k),x(t_k))$ represents the value of the cost function at the effectively {\em visited} pairs $(\hat p(t_k),x(t_k))$. Therefore, the difference expressed in the l.h.s of (\ref{lhs45}) is relevant for the stability assessment of the resulted truncated MPC implementation. On the other hand, using the definition (\ref{defdetauk}) of $\tau_k$, the r.h.s of (\ref{lhs45}) can be viewed as a function of the precision pair $(\varepsilon_0^{(k+1)},\varepsilon_\psi^{(k+1)})$. The stability issue is therefore dependent on the possibility to define these precision parameters in such a way that the r.h.s of (\ref{lhs45}) is negative. This is the aim of the following development. \ \\ \ \\ 
Since the only negative term in the r.h.s of (\ref{lhs45}) is $-\Delta(\tau_k,x(t_k))$, we need a lower bound on $\Delta(\tau_k,x(t_k))$. The following straightforward lemma gives such a lower bound:\\ 
\begin{lemma} \label{simplergloballemma} 
{\bf If}  the following conditions hold:
\begin{enumerate}
\item $(\hat p^*(t_k),x(t_k))\in \mathbb C$
\item Assumption \ref{asslC} is satisfied
\end{enumerate} 
{\bf then} a computable lower bound of the quantity $\Delta(\tau,x(t_k))$ can be obtained by:
\begin{eqnarray}
\Delta(\tau,x(t_k))\ge \Gamma_\mathbb{C}(\tau,q(x(t_k)) \label{PLdSSQ} 
\end{eqnarray} 
where $\Gamma_\mathbb{C}(\tau,q)$ is given by (see Figure \ref{figGamma}): 
\begin{eqnarray}
\Gamma_\mathbb{C}(\tau,q):=\left\{ 
\begin{array}{ll}
q\tau-\dfrac{1}{2}D_{\mathbb C}\tau^2 & \mbox{\rm if $\tau\le q/D_{\mathbb C}$}\\
\dfrac{q^2}{2D_{\mathbb C}} & \mbox{otherwise}
\end{array}\right. \label{gf52MLOK} 
\end{eqnarray}
\end{lemma}
{\sc Proof}. See Appendix \ref{proofsimplergloballemma}.  \ \\ \ \\
\noindent Using the definition (\ref{defdetauk}) of $\tau_k$ and the r.h.s of (\ref{PLdSSQ}) in (\ref{lhs45}) the following computable function can be defined:
\begin{eqnarray}
&&R_{\tau_c}(\varepsilon_0,\varepsilon_\psi,\bar q):=K^0_\mathbb{C}(E^0_\mathbb{C}+\tau_cE^1_\mathbb{C}\bar N(\varepsilon_0,\varepsilon_\psi))+\varepsilon_0-\nonumber \\
&&\Gamma_{\mathbb C}(\tau_c\cdot \bar N(\varepsilon_0,\varepsilon_\psi),\bar q)  \label{defdeRR} 
\end{eqnarray} 
so that the following corollary of Lemma \ref{lemfondineq} can be stated:\\
\begin{corollary} \label{corcorcor} 
{\bf If} the following conditions hold
\begin{enumerate}
\item The requirements of Lemma \ref{lemfondineq}  are satisfied
\item Assumption \ref{asslC} holds
\item $q(x(t_k))\ge \bar q$
\end{enumerate} 
{\bf then} the following inequality holds
\begin{eqnarray}
&&f_0(\hat p^*(t_{k+1}),x(t_{k+1}))-f_0(\hat p^*(t_{k}),x(t_{k}))\le \nonumber \\
&&\varepsilon_0^{(k)}+R_{\tau_c}(\varepsilon_0^{(k+1)},\varepsilon_\psi^{(k+1)},\bar q)\label{lhs45rewritten2} 
\end{eqnarray} 
where $R_{\tau_c}(\cdot)$ is defined by (\ref{defdeRR}).\\
\end{corollary}
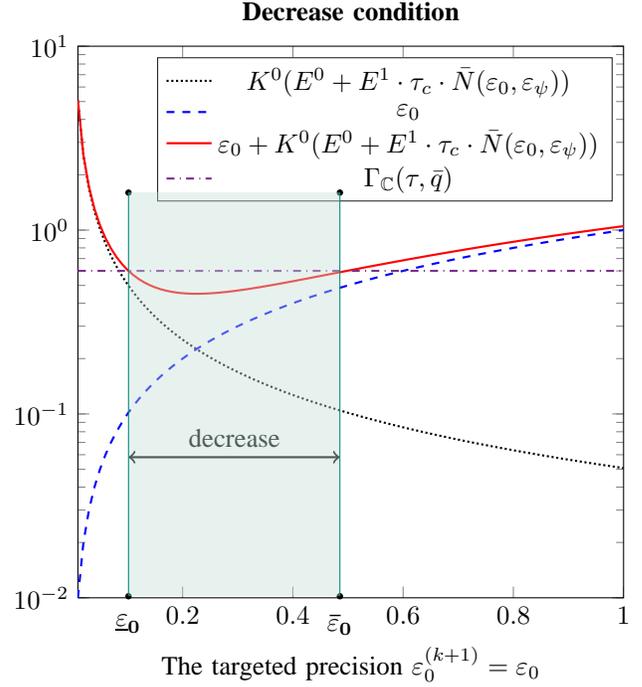
\begin{figure}
\begin{center}
%

\tikzstyle{dashdotted}=              [dash pattern=on 3pt off 2pt on \the\pgflinewidth off 2pt]
\definecolor{mycolor1}{rgb}{0.00000,0.44700,0.74100}%
\definecolor{mycolor2}{rgb}{0.85000,0.32500,0.09800}%
\definecolor{mycolor3}{rgb}{0.92900,0.69400,0.12500}%
\definecolor{mycolor4}{rgb}{0.49400,0.18400,0.55600}%
\begin{tikzpicture}

\begin{axis}[%
width=0.4\textwidth,
height=0.3\textheight,
at={(1.011111in,0.641667in)},
scale only axis,
xmin=0.01,
xmax=1,
xlabel={The targeted precision $\varepsilon_0^{(k+1)}=\varepsilon_0$},
ymode=log,
ymin=0.01,
ymax=10,
title style={font=\bfseries},
title={Decrease condition},
legend style={draw=white!15!black}
]
\addplot [color=black,densely dotted,thick]
  table[row sep=crcr]{%
0.01	5.06837703975989\\
0.02	2.53423820365218\\
0.03	1.68952525828294\\
0.04	1.26716878559832\\
0.05	1.01375490198755\\
0.06	0.844812312913704\\
0.07	0.724139035003814\\
0.08	0.633634076571395\\
0.09	0.563241331123959\\
0.1	0.50692713476601\\
0.11	0.460851883200415\\
0.12	0.422455840229086\\
0.13	0.389966880791808\\
0.14	0.362119201274141\\
0.15	0.337984545692163\\
0.16	0.316866722057932\\
0.17	0.298233348263022\\
0.18	0.281670349334213\\
0.19	0.266850823976858\\
0.2	0.253513251155239\\
0.21	0.24144592336425\\
0.22	0.230475625372442\\
0.23	0.220459266336443\\
0.24	0.211277603886777\\
0.25	0.202830474433085\\
0.26	0.195033124168138\\
0.27	0.187813355404298\\
0.28	0.181109284409304\\
0.29	0.174867563138103\\
0.3	0.169041956618315\\
0.31	0.163592195680449\\
0.32	0.1584830448012\\
0.33	0.153683539429784\\
0.34	0.149166357903745\\
0.35	0.144907301036337\\
0.36	0.140884858439341\\
0.37	0.137079845171912\\
0.38	0.133475095760663\\
0.39	0.130055205293581\\
0.4	0.126806309349853\\
0.41	0.123715896135088\\
0.42	0.120772645454359\\
0.43	0.117966290154129\\
0.44	0.115287496458455\\
0.45	0.112727760260366\\
0.46	0.110279316940455\\
0.47	0.107935062697988\\
0.48	0.105688485715623\\
0.49	0.103533605752946\\
0.5	0.101464920988776\\
0.51	0.0994773611173193\\
0.52	0.0975662458563029\\
0.53	0.0957272481523061\\
0.54	0.0939563614743831\\
0.55	0.0922498706756574\\
0.56	0.0906043259768862\\
0.57	0.0890165196885981\\
0.58	0.0874834653412855\\
0.59	0.0860023789379497\\
0.6	0.0845706620813916\\
0.61	0.0831858867611142\\
0.62	0.0818457816124586\\
0.63	0.0805482194843953\\
0.64	0.0792912061728339\\
0.65	0.078072870193936\\
0.66	0.0768914534871258\\
0.67	0.0757453029506683\\
0.68	0.0746328627241065\\
0.69	0.0735526671417929\\
0.7	0.0725033342904025\\
0.71	0.0714835601108823\\
0.72	0.0704921129919043\\
0.73	0.0695278288076929\\
0.74	0.0685896063581899\\
0.75	0.0676764031740069\\
0.76	0.0667872316525656\\
0.77	0.0659211554953176\\
0.78	0.0650772864190246\\
0.79	0.0642547811168151\\
0.8	0.0634528384471607\\
0.81	0.0626706968310781\\
0.82	0.061907631839778\\
0.83	0.0611629539567019\\
0.84	0.0604360064994134\\
0.85	0.0597261636881788\\
0.86	0.0590328288492984\\
0.87	0.0583554327423464\\
0.88	0.0576934320014614\\
0.89	0.0570463076817199\\
0.9	0.0564135639024171\\
0.91	0.0557947265798023\\
0.92	0.0551893422424617\\
0.93	0.0545969769231284\\
0.94	0.0540172151212277\\
0.95	0.0534496588309461\\
0.96	0.0528939266300453\\
0.97	0.0523496528250393\\
0.98	0.0518164866487069\\
0.99	0.0512940915062399\\
1	0.0507821442666222\\
};
\addlegendentry{$K^0(E^0+E^1\cdot\tau_c\cdot\bar N(\varepsilon_0,\varepsilon_\psi))$};

\addplot [color=blue,dashed,thick]
  table[row sep=crcr]{%
0.01	0.01\\
0.02	0.02\\
0.03	0.03\\
0.04	0.04\\
0.05	0.05\\
0.06	0.06\\
0.07	0.07\\
0.08	0.08\\
0.09	0.09\\
0.1	0.1\\
0.11	0.11\\
0.12	0.12\\
0.13	0.13\\
0.14	0.14\\
0.15	0.15\\
0.16	0.16\\
0.17	0.17\\
0.18	0.18\\
0.19	0.19\\
0.2	0.2\\
0.21	0.21\\
0.22	0.22\\
0.23	0.23\\
0.24	0.24\\
0.25	0.25\\
0.26	0.26\\
0.27	0.27\\
0.28	0.28\\
0.29	0.29\\
0.3	0.3\\
0.31	0.31\\
0.32	0.32\\
0.33	0.33\\
0.34	0.34\\
0.35	0.35\\
0.36	0.36\\
0.37	0.37\\
0.38	0.38\\
0.39	0.39\\
0.4	0.4\\
0.41	0.41\\
0.42	0.42\\
0.43	0.43\\
0.44	0.44\\
0.45	0.45\\
0.46	0.46\\
0.47	0.47\\
0.48	0.48\\
0.49	0.49\\
0.5	0.5\\
0.51	0.51\\
0.52	0.52\\
0.53	0.53\\
0.54	0.54\\
0.55	0.55\\
0.56	0.56\\
0.57	0.57\\
0.58	0.58\\
0.59	0.59\\
0.6	0.6\\
0.61	0.61\\
0.62	0.62\\
0.63	0.63\\
0.64	0.64\\
0.65	0.65\\
0.66	0.66\\
0.67	0.67\\
0.68	0.68\\
0.69	0.69\\
0.7	0.7\\
0.71	0.71\\
0.72	0.72\\
0.73	0.73\\
0.74	0.74\\
0.75	0.75\\
0.76	0.76\\
0.77	0.77\\
0.78	0.78\\
0.79	0.79\\
0.8	0.8\\
0.81	0.81\\
0.82	0.82\\
0.83	0.83\\
0.84	0.84\\
0.85	0.85\\
0.86	0.86\\
0.87	0.87\\
0.88	0.88\\
0.89	0.89\\
0.9	0.9\\
0.91	0.91\\
0.92	0.92\\
0.93	0.93\\
0.94	0.94\\
0.95	0.95\\
0.96	0.96\\
0.97	0.97\\
0.98	0.98\\
0.99	0.99\\
1	1\\
};
\addlegendentry{$\varepsilon_0$};

\addplot [color=red,solid,thick]
  table[row sep=crcr]{%
0.01	5.07837703975989\\
0.02	2.55423820365218\\
0.03	1.71952525828294\\
0.04	1.30716878559832\\
0.05	1.06375490198755\\
0.06	0.904812312913704\\
0.07	0.794139035003814\\
0.08	0.713634076571395\\
0.09	0.653241331123959\\
0.1	0.60692713476601\\
0.11	0.570851883200415\\
0.12	0.542455840229086\\
0.13	0.519966880791808\\
0.14	0.502119201274141\\
0.15	0.487984545692163\\
0.16	0.476866722057932\\
0.17	0.468233348263022\\
0.18	0.461670349334213\\
0.19	0.456850823976858\\
0.2	0.453513251155239\\
0.21	0.45144592336425\\
0.22	0.450475625372442\\
0.23	0.450459266336443\\
0.24	0.451277603886777\\
0.25	0.452830474433085\\
0.26	0.455033124168138\\
0.27	0.457813355404298\\
0.28	0.461109284409304\\
0.29	0.464867563138103\\
0.3	0.469041956618315\\
0.31	0.473592195680449\\
0.32	0.4784830448012\\
0.33	0.483683539429784\\
0.34	0.489166357903745\\
0.35	0.494907301036337\\
0.36	0.500884858439341\\
0.37	0.507079845171912\\
0.38	0.513475095760663\\
0.39	0.520055205293581\\
0.4	0.526806309349853\\
0.41	0.533715896135088\\
0.42	0.540772645454359\\
0.43	0.547966290154129\\
0.44	0.555287496458455\\
0.45	0.562727760260366\\
0.46	0.570279316940455\\
0.47	0.577935062697988\\
0.48	0.585688485715623\\
0.49	0.593533605752946\\
0.5	0.601464920988776\\
0.51	0.609477361117319\\
0.52	0.617566245856303\\
0.53	0.625727248152306\\
0.54	0.633956361474383\\
0.55	0.642249870675657\\
0.56	0.650604325976886\\
0.57	0.659016519688598\\
0.58	0.667483465341286\\
0.59	0.67600237893795\\
0.6	0.684570662081392\\
0.61	0.693185886761114\\
0.62	0.701845781612459\\
0.63	0.710548219484395\\
0.64	0.719291206172834\\
0.65	0.728072870193936\\
0.66	0.736891453487126\\
0.67	0.745745302950668\\
0.68	0.754632862724106\\
0.69	0.763552667141793\\
0.7	0.772503334290403\\
0.71	0.781483560110882\\
0.72	0.790492112991904\\
0.73	0.799527828807693\\
0.74	0.80858960635819\\
0.75	0.817676403174007\\
0.76	0.826787231652566\\
0.77	0.835921155495318\\
0.78	0.845077286419025\\
0.79	0.854254781116815\\
0.8	0.863452838447161\\
0.81	0.872670696831078\\
0.82	0.881907631839778\\
0.83	0.891162953956702\\
0.84	0.900436006499413\\
0.85	0.909726163688179\\
0.86	0.919032828849298\\
0.87	0.928355432742346\\
0.88	0.937693432001461\\
0.89	0.94704630768172\\
0.9	0.956413563902417\\
0.91	0.965794726579802\\
0.92	0.975189342242462\\
0.93	0.984596976923128\\
0.94	0.994017215121228\\
0.95	1.00344965883095\\
0.96	1.01289392663005\\
0.97	1.02234965282504\\
0.98	1.03181648664871\\
0.99	1.04129409150624\\
1	1.05078214426662\\
};
\addlegendentry{$\varepsilon_0+K^0(E^0+E^1\cdot\tau_c\cdot\bar N(\varepsilon_0,\varepsilon_\psi))$};

\addplot [color=mycolor4,dashdotted,thick]
  table[row sep=crcr]{%
0.01	0.6\\
0.02	0.6\\
0.03	0.6\\
0.04	0.6\\
0.05	0.6\\
0.06	0.6\\
0.07	0.6\\
0.08	0.6\\
0.09	0.6\\
0.1	0.6\\
0.11	0.6\\
0.12	0.6\\
0.13	0.6\\
0.14	0.6\\
0.15	0.6\\
0.16	0.6\\
0.17	0.6\\
0.18	0.6\\
0.19	0.6\\
0.2	0.6\\
0.21	0.6\\
0.22	0.6\\
0.23	0.6\\
0.24	0.6\\
0.25	0.6\\
0.26	0.6\\
0.27	0.6\\
0.28	0.6\\
0.29	0.6\\
0.3	0.6\\
0.31	0.6\\
0.32	0.6\\
0.33	0.6\\
0.34	0.6\\
0.35	0.6\\
0.36	0.6\\
0.37	0.6\\
0.38	0.6\\
0.39	0.6\\
0.4	0.6\\
0.41	0.6\\
0.42	0.6\\
0.43	0.6\\
0.44	0.6\\
0.45	0.6\\
0.46	0.6\\
0.47	0.6\\
0.48	0.6\\
0.49	0.6\\
0.5	0.6\\
0.51	0.6\\
0.52	0.6\\
0.53	0.6\\
0.54	0.6\\
0.55	0.6\\
0.56	0.6\\
0.57	0.6\\
0.58	0.6\\
0.59	0.6\\
0.6	0.6\\
0.61	0.6\\
0.62	0.6\\
0.63	0.6\\
0.64	0.6\\
0.65	0.6\\
0.66	0.6\\
0.67	0.6\\
0.68	0.6\\
0.69	0.6\\
0.7	0.6\\
0.71	0.6\\
0.72	0.6\\
0.73	0.6\\
0.74	0.6\\
0.75	0.6\\
0.76	0.6\\
0.77	0.6\\
0.78	0.6\\
0.79	0.6\\
0.8	0.6\\
0.81	0.6\\
0.82	0.6\\
0.83	0.6\\
0.84	0.6\\
0.85	0.6\\
0.86	0.6\\
0.87	0.6\\
0.88	0.6\\
0.89	0.6\\
0.9	0.6\\
0.91	0.6\\
0.92	0.6\\
0.93	0.6\\
0.94	0.6\\
0.95	0.6\\
0.96	0.6\\
0.97	0.6\\
0.98	0.6\\
0.99	0.6\\
1	0.6\\
};
\addlegendentry{$\Gamma_\mathbb{C}(\tau,\bar q)$};
\end{axis}
\draw[fill=black] (3.24,7.02) node (p1){} circle (1pt);
\draw[fill=black] (6.05,7.02) node (p2){} circle (1pt);
\draw[fill=black] (6.05,1.65) node (p3){} circle (1pt);
\draw[fill=black] (3.24,1.65) node (p4){} circle (1pt);
\node[below] at(p4.south) {$\mathbf{\underline \varepsilon_0}$};
\node[below] at(p3.south) {$\mathbf{\bar \varepsilon_0}$};
\draw[<->,thick] (3.24,3.5) -- node[midway,above]{decrease} (6.05,3.5);
\fill[fill=PineGreen!20,opacity=0.4] (p1.base) -- (p2.base) -- (p3.base) -- (p4.base) -- cycle;
\draw[PineGreen] (p1.base) -- (p4.base);
\draw[PineGreen] (p2.base) -- (p3.base);
\end{tikzpicture}%
\end{center} 
\caption{\color{Blue} Typical evolution of the quantities involved in the r.h.s of equation (\ref{lhs45rewritten2}) invoked in corollary \ref{corcorcor}. The decrease of the cost function is possible if there is a tagreted future precision $\varepsilon_0$ for which the red-solid curve lies below the dash-dotted curve.}\label{figdecreasecondition} 
\end{figure}
Figure \ref{figdecreasecondition} presents a typical situation showing that for a given past achieved precision $\varepsilon_0^{(k)}$, a given computational power leading to the computation time $\tau_c$ and a given precision $\varepsilon_\psi$ on the soft constraints satisfaction, either there is no $\varepsilon_0^{(k+1)}$ making the r.h.s of  equation (\ref{defdeRR}) invoked in corollary \ref{corcorcor} negative or there is an interval of successful values of $\varepsilon_0^{(k+1)}$ which does not contain $0$ and which depends on the current value of $q(x(t_k))=\bar q$. \ \\ \ \\ 
Note that corollary \ref{corcorcor} involves quantities that depend on some compact set to which belong all the pair $([\hat p^*(t_k)]^{+\tau_k},\hat x(t_{k+1})$. Using assumption \ref{assCphi}, it is possible to prove that such compact set is linked to a set of initial conditions for which a certified convergence result can be derived for the resulting real-time MPC. This is stated in the following proposition which is the main contribution of the paper:\\
\begin{propos}\label{enfinpropos} 
Consider a positive real $\phi_0>0$ and the corresponding compact subset $\mathbb C_{\phi_0}\subset \mathbb R^{n_p}\times \mathbb{R}^{n}$ defined according to assumption \ref{assCphi}. Let be given a precision $\varepsilon_\psi>0$ on the soft constraints satisfaction. \ \\ \ \\ {\bf If} the following conditions hold with $\mathbb C=\mathbb C_{\phi_0}$:
\begin{enumerate}
\item Assumptions \ref{assprederror}, \ref{assdx}, \ref{assDelta} and \ref{asslC} are satisfied
\item $\exists\bar q_{min}>0$ and $\gamma_c>0$ such that the inequality:
\begin{eqnarray}
R_{\tau_c}(\varepsilon_0,\varepsilon_\psi,\bar q)\le -\left[\dfrac{\gamma_c\bar q_{min}^2}{3D_{\mathbb C_{\phi_0}}}\right] \label{tgf56} 
\end{eqnarray}
admits a solution $\varepsilon_0^{sol}(\bar q)\in [0,\gamma_c\bar q_{min}^2/(2D_{\mathbb C_{\phi_0}})]$ for all $\bar q\ge \bar q_{min}$
\end{enumerate}  
{\bf then} the truncated MPC design based on the adaptive sampling period defined by:
\begin{eqnarray}
\tau_k:=\tau_c\times \bar N(\varepsilon_0^{sol}(q(x(t_k)),\varepsilon_\psi) \label{sdsam} 
\end{eqnarray} 
steers the system to the set:
\begin{eqnarray}
\mathbb X_{min}:=\Bigl\{x\in \mathbb{R}^{n}\quad\vert\quad q(x)\le \bar q_{min}\Bigr\} \label{defdeXmin} 
\end{eqnarray} 
provided that the initial condition satisfies:
\begin{eqnarray}
f_0(\hat p^*(t_0),x(t_0))\le \phi_0\quad;\quad \varepsilon_0^{(0)}\le \dfrac{\gamma_c\bar q_{min}^2}{6D_{\mathbb C_{\phi_0}}} \label{jhghghjg} 
\end{eqnarray} 
{\bf Moreover}, if the hard constraints depend only on $p$, then along the closed-loop trajectory, one has:
\begin{eqnarray}
\max_{i\in I_h, k\ge 0} [c_i(\hat p^*(t_k),x(t_k))]&\le& 0\nonumber \\
\max_{i\in I_s, k\ge 0} [c_i(\hat p^*(t_k),x(t_k))]&\le& \varepsilon_\psi+\label{viuolatesfot}   \\ &&+K_{\mathbb C_{\phi_0}}^\psi\cdot(E^0_{\mathbb C_{\phi_0}}++E^1_{\mathbb C_{\phi_0}}\tau_k)\nonumber 
\end{eqnarray} 
\end{propos}
{\sc Proof}. See Appendix \ref{proofenfinpropos}.   
\subsection{Case of linear MPC} \label{QPMPCCCC} 
\noindent Linear MPC formulation applies to system of the form
\begin{eqnarray}
\dot z=A_0 z+B_0 u \label{physical1} 
\end{eqnarray} 
in order to stabilize the physical state $z$ around some desired value $z_d$. We assume for the sake of simplicity that $z_d$ is a steady state for (\ref{physical1}) that corresponds to the steady control $u_d=0$. Using the extended system with the extended state $x=(z^T,z_d^T)$ and the extended dynamic built up using (\ref{physical1}) with $\dot z_d=0$, one obtains the controlled system model given by: 
\begin{eqnarray}
\dot x=A_sx+B_su
\end{eqnarray} 
where $x$ is an extended state containing the set-point and disturbance model state and where the cost function (\ref{de548OI}) is given by:
\begin{eqnarray}
\bar\ell(s,p,x):=\dfrac{1}{2}\bigl[q(\bar x(s,p,x))+\|\mathcal U(s,p)\|_R^2\bigr] \label{costLMPC} 
\end{eqnarray} 
where $q(x)$ is given by:
\begin{eqnarray}
q(x)=\|z-z_d\|_Q^2:= \|Cx\|_Q^2 \label{ghgFFFF56} 
\end{eqnarray} 
The control parametrization map $\mathcal U(\cdot,p)$ used in (\ref{costLMPC}) gives the control profile over the prediction horizon as a function of the finite dimensional parameter vector $p$. \ \\ \ \\ 
This formulation leads to state-dependent QP where the cost function and the constraints are given by:
\begin{eqnarray}
&&f_0(p,x)=\dfrac{1}{2}p^THp+(F_1x)^Tp+x^TSx \label{costLMPC}\\
&&Ap\le B^{(0)}+B^{(1)}x  \label{constraintsLMPC} 
\end{eqnarray}
It results that the definition of $L_0$, $L_\psi$ and $\mu_0$ remains unchanged since these parameters depends only on the state independent quantities $H$ and $A$. \ \\ \ \\ It is also assumed that the formulation involves appropriate final constraints such that (\ref{gfTTRTRDE}) of Assumption \ref{assDelta} holds with $\Delta(\tau,x)$ satisfying: 
\begin{eqnarray}
\Delta(\tau,x)\ge \int_0^\tau q(\bar x(s,p^{opt},x))ds
\end{eqnarray} 
This can be obtained through appropriate final equality constraints that can be explicitly embedded in the control parametrization map $\mathcal U(\cdot,p)$ or through softened final inequality constraints as suggested in \cite{Mayne:2000}. \ \\ \ \\ 
Given a set of interest $\mathbb X$, the upper bound on $D_0$ defined by (\ref{rftdrefRR})-(\ref{defdefbar}) can be used provided that $\|F\|$ is replaced by 
\begin{eqnarray}
\sup_{x\in \mathbb X} \|F_1x\| \le  \|F_1\|\times \varrho(\mathbb X)\label{defdeFFF1}
\end{eqnarray} 
The computation of $\psi^{max}$ invoked in (\ref{defdepsimaxhaut}) of proposition \ref{ygtf7865} is obtained according to: 
\begin{eqnarray}
\psi^{max}:=\max_{x\in \mathbb X}\left[\sum_{i=1}^{n_c}\left(\max\bigl\{0,M_ix-L_i\bigr\}\right)^2\right] \label{defdepsimaxxxx} 
\end{eqnarray} 
where 
\begin{eqnarray*}
M_i&:=&-\left[A_iH^{-1}F_1+B^{(1)}_i\right] \label{defdeMi} \\
L_i&:=&B_i^{(0)} \label{defdeLi} 
\end{eqnarray*} 
where $A_i$ and $B_i^{(j)}$ denote the $i$-th line of $A$ and $B^{(j)}$ respectively. Note that the optimization problems (\ref{defdepsimaxxxx}) can be computed once for all using available NLP solvers for a beforehand given sets of interest $\mathbb P$ and $\mathbb X$. \ \\ \ \\ Once $\psi^{max}$ is computed, the resulting $\kappa_0^{max}$ involved in Proposition \ref{ygtf7865} [item (4)] can be computed and used in the computation of $\rho^{max}$. Finally, the parameter $f_0^{max}$ involved in Proposition \ref{ygtf7865} is computed according to:
\begin{eqnarray*}
f_0^{max}:=\max_{(p,x)\in \mathbb P\times \mathbb X}\left[ \begin{pmatrix}p \cr x
\end{pmatrix}^T  \underbrace{\begin{pmatrix}
H&F_1\cr F_1^T&S
\end{pmatrix}}_{:=W} \begin{pmatrix}
p\cr x
\end{pmatrix}  \right]  \label{defdef0maxxxx} 
\end{eqnarray*}  
which admits the upper bound:
\begin{eqnarray*}
f_0^{max}&\le& \left[\lambda_{max}(W)\right]\times \max_{z\in \mathbb P\times \mathbb X} \|z\|^2 \\
&=&\lambda_{max}(W)\times \varrho(\mathbb P\times \mathbb X)
\end{eqnarray*} 
It remains to give explicit computation of $D_{\mathbb C}$ in (\ref{qx}) of Assumption \ref{asslC}. This is given by the following proposition: \\
\begin{propos}
If the constraints $p\in \mathbb P$ implies that $\mathcal U(s,p)\in \mathbb U$ for some compact set $\mathbb U$, then the following expression of $D_{\mathbb C}$ meets the requirement of Assumption \ref{asslC}:
\begin{eqnarray}
D_{\mathbb C}=\lambda_{max}(Q)\times \varrho(\mathbb X)\times \left[\|A_s\|\varrho(\mathbb X)+\|B_s\|\varrho(\mathbb U)\right]
\end{eqnarray} 
\end{propos}
{\sc Proof}. Compute the derivative of $\|\bar Cx(s,p,x)\|_Q^2$ (which takes the values $q(x)$ at $s=0$) and derive a lower bound on the speed with which this term may converge to $0$ given the compact set to which belongs the arguments $x$ and $p$. $\hfill \Box$ \ \\ \ \\ 
The next result concerns the explicit derivation of the compact set $\mathbb C_{\phi}$ given an initial cost function level $\phi$ as described in Assumption \ref{assCphi}. This is the aim of the following result:
\begin{propos} \label{proposphi0Cphi0} 
{\bf If} The possible set points $z_d$ belong to a compact set $\mathbb Z_d$, 
{\bf then} given $\phi$, the compact set $\mathbb C_\phi$ involved in (\ref{defdeCphibarbarbar}) of Assumption \ref{assCphi}  is given by  $\mathbb C_\phi:=\mathbb P\times \mathbb X$ where:
\begin{eqnarray}
\mathbb P&:=&\left\{p \quad \mbox{\rm s.t} \quad \|p\|\le \left[\phi/\lambda_{min}(W_0)\right]^{\frac{1}{2}}\right\} \label{defdePbb} \\
\mathbb X&:=&\left\{x \quad \mbox{\rm s.t} \quad \|x\|\le \varrho(\mathbb Z_d)+\left[\phi/\lambda_{min}(W_0)\right]^{\frac{1}{2}}\right\} \label{gfGGvF} 
\end{eqnarray} 
where $W_0$ is the matrix given by:
\begin{eqnarray}
W_0=\begin{pmatrix}
H&F_{11}\cr F_{11}^T&S_{11}
\end{pmatrix}
\end{eqnarray}  
where $F_{11}\in \mathbb{R}^{n_p\times n_z}$ and $S_{11}\in \mathbb{R}^{n_z\times n_z}$ are the $z$-corresponding sub-matrices of $F_1$ and $S$ involved in (\ref{costLMPC}) respectively.  \\
\end{propos}
{\sc Proof} See Appendix \ref{proofproposphi0Cphi0}. \ \\ \ \\ 
Once the compact set $\mathbb C_\phi$ is computed for a given initial value $\phi$ of the cost function, the constants $K_{\mathbb C}^0$ and $K_{\mathbb C}^1$ involved in (\ref{defdeKdep}) and (\ref{defdeKdepsi})  of Assumption \ref{assdx} can be explicitly computed using the following proposition:\\
\begin{propos} \label{propossssss76} 
For the cost function $f_0$ defined by (\ref{costLMPC}) and the constraints defined by (\ref{constraintsLMPC}), given a compact set $\mathbb C:=\mathbb P\times \mathbb X$, the constants  $K_{\mathbb C}^0$ and $K_{\mathbb C}^1$ involved in (\ref{defdeKdep}) and (\ref{defdeKdepsi})  of Assumption \ref{assdx} can be given by:
\begin{eqnarray}
K_\mathbb{C}^0&:=& \|F_1^T\|\times \varrho(\mathbb P)+2\lambda_{max}(S)\times \varrho(\mathbb X)\label{expressionKC0}\\
K_{\mathbb C}^1&:=& 2n_c\left[\psi^{max}\right]\cdot \|(B^{(1)})^T\| \label{expressionKC1} 
\end{eqnarray} 
where $\psi^{max}$ is computed by (\ref{defdepsimaxxxx}).  \\
\end{propos}
{\sc Proof}. See Appendix \ref{proofpropossssss76}.  \ \\ \ \\ 
The only remaining parameters are $E_\mathbb{C}^0$ and $E_\mathbb{C}^1$ involved in (\ref{defdeE0E1}) of Assumption \ref{assprederror} and which describe the prediction error on the extended state $x$ as a function of $\tau$. Note that if the model is perfectly known, the only prediction error comes from the fact that the future evolution of the set-point $z_d$ is unknown. Two cases can be distinguished:
\begin{itemize}
\item If the set point if filtered, then 
\begin{eqnarray}
\mbox{\rm $E_\mathbb{C}^0=0$ and $E_\mathbb{C}^1=\max_t(\|\dot z_d(t)\|)$} \label{E0E11}
\end{eqnarray} 
\item Otherwise 
\begin{eqnarray}
\mbox{\rm $E_\mathbb{C}^0=\varrho(\mathbb Z_d)$ and $E_\mathbb{C}^1=\max_t(\|\dot z_d(t)\|)$} \label{E0E12}  
\end{eqnarray} 
\end{itemize} 
In case other sources of prediction errors prevail, then an additional positive term $e_1$ has to be added so that $E_\mathbb{C}^1=\max_t(\|\dot z_d(t)\|)+e_1$ is used. \\
\subsubsection{Illustrative example: MPC control of a chain of integrators}
\begin{figure}
\begin{center}
\input{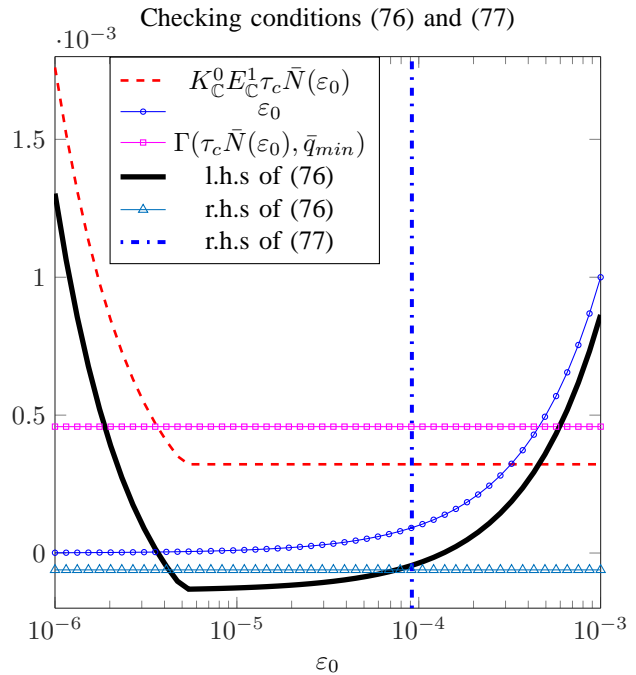}
\end{center} 
\caption{Check of the certification feasibility for the chain of $n=4$ integrators. The condition (\ref{hg87650}) is satisfied by an interval of values of $\varepsilon_0$ including values satisfying (\ref{GGGG}). Successful values: $\bar q_{min}=0.36$, $\gamma_c=0.2$, $E_\mathbb{C}^1=0.05$.} \label{cas41} 
\end{figure}

Let us consider MPC control of a chain of $n$ integrators given by:
\begin{eqnarray}
\dot z_i&=&z_{i+1}\quad \mbox{\rm for $i=1,\dots,n-1$}\\
\dot z_n&=&u \quad \mbox{\rm under $\vert u\vert\le  \bar u=10$} \label{gftrfr09} 
\end{eqnarray}  
in which the objective is to track a reference trajectory on $z_1$ under the state constraints:
\begin{eqnarray}
\begin{pmatrix}
-2\cr -1
\end{pmatrix} \le \begin{pmatrix}
z_1(t)\cr z_2(t)
\end{pmatrix} \le \begin{pmatrix}
+2\cr +1
\end{pmatrix} 
\end{eqnarray} 
using the formulation of section \ref{QPMPCCCC}. This is obviously a very important sub-class of systems that is heavily used in Mechatronics. \ \\ \ \\ 
We consider a parametrization of the form:
\begin{eqnarray}
\mathcal U(s,p_u(t))=\left[\Phi_u(s)\right]p_u(t)\quad \quad;\quad \mbox{\rm $p_u\in \mathbb{R}^{m}$} \label{gfRTF4} 
\end{eqnarray}  
in which a final constraint on the state is imposed:
\begin{eqnarray}
 \|z(T)-Z_d\|=Cx(T)=0 \label{finalcccc} 
\end{eqnarray} By doing this, the stability of the ideal perfect scheme is guaranteed with Assumption \ref{assDelta} satisfied. The final constraint satisfaction can be imposed through the reduced parametrization:
\begin{eqnarray}
p_u=Kp+Mx_0 \label{defdeKetM} 
\end{eqnarray}
where the matrices $K$ and $M$ depend on the function basis $\phi_u$ involved in (\ref{gfRTF4}) and the prediction horizon $T$ such that taking $p=0$ always leads to $p_u$ that satisfies the final constraint. 
This means that because of the saturation constraint (\ref{gftrfr09}), the final constraint (\ref{finalcccc})  can be feasible through (\ref{defdeKetM}) only for initial state $x_0$ such that 
\begin{eqnarray}
\|Mx_0\|\le \bar u \label{defdepu} 
\end{eqnarray} 
leading to the bound $\|x_0\|\le \bar u/\|M\|$ ($\approx 11.3$ in the case $n=4$). By doing this, the number of decision variables is given by $n_p=m-n$. In the following results, the weighting matrices $Q=\mathbb I_n$ and $R=0.001$ are systematically used in (\ref{ghgFFFF56}) and (\ref{costLMPC}). A prediction horizon $T=10$ is used in the sequel while $m=10$ dimensional parametrization variable is used in control parametrization (\ref{gfRTF4}). This leads to a number of free decision variables of dimension $n_p=6$.  \ \\ \ \\ 
The computation time for a single iteration $\tau_c=0.1 \mu s$ is used (time needed for a matrix-vector multiplication in Step 10 of Algorithm 2). It is supposed that the reference value $z_d$ lies in the domain $[-5,+5]$. All the constraints are taken to be soft with $\varepsilon_\psi=10^{-2}$. The formulation of the problem and the choice of the constraints checking instants lead to a number of constraints $n_c=300$. \ \\ \ \\ 
Note that in order to check the existence of $q_{min}$ satisfying the condition (\ref{tgf56}) of Proposition (\ref{enfinpropos}), one can check the existence of solution $\varepsilon_0^*$ to the inequality 
\begin{eqnarray}
R_{\tau_c}(\varepsilon_0,\varepsilon_\psi,\bar q_{min})\le -\left[\dfrac{\gamma_c\bar q_{min}^2}{3D_{\mathbb C_{\phi_0}}}\right] \label{hg87650} 
\end{eqnarray} 
since if (\ref{hg87650}) is satisfied for $\bar q_{min}$ in the l.h.s, it will be satisfied for any $\bar q\ge \bar q_{min}$ used in the l.h.s while $\bar q_{min}$ is used in the r.h.s. An additional condition invoked in Proposition \ref{enfinpropos} states that this solution $\varepsilon_0^*$ must be such that:
\begin{eqnarray}
\varepsilon_0^*\le \left[\dfrac{\gamma_c\bar q_{min}^2}{2D_{\mathbb C_{\phi_0}}}\right] \label{GGGG} 
\end{eqnarray} 
Note also that the parameter $\phi_0$ invoked in (\ref{hg87650}) defines an upper bound on the possible initial value of the cost $f_0(p,x)$. Therefore, in the case of hot starts, the size of $\phi_0$ can define the quality of the hot start. Otherwise, one can take an upper bound pessimistic value $\phi_0$ by starting from $p=0$ and taking the upper value of $f_0(0,x)$ over the set of admissible initial state defined by (\ref{defdepu}), namely:
\begin{eqnarray}
\phi_0\le \lambda_{max}(S)\dfrac{\bar u}{\|M\|} \label{defdephiphi0} 
\end{eqnarray} 
Based on the knowledge of $\phi_0$ given by (\ref{defdephiphi0}), the condition (\ref{hg87650}) can be checked for different candidate values of $\bar q_{min}$.\ \\ \ \\ 
Figures \ref{cas41} and \ref{cas21} shows the results for the cases $n=4$ and $n=2$ respectively. More precisely, Figure \ref{cas41} shows that for the quadruple integrator system under an unknown future behavior of the set-point characterized by $E_\mathbb{C}^1=0.05$, the certification conditions (\ref{hg87650}) and (\ref{GGGG}) are satisfied with $\bar q_{min}=0.36$ and $\gamma_c=0.2$. \ \\ \ \\ Figure \ref{cas21} shows that the certification is possible for the double integrator system with the unknown behavior of the set-point defined by $E_\mathbb{C}^1=0.3$ provided that $\bar q_{min}=0.18$ and $\gamma_c$ are used in (\ref{hg87650}) and (\ref{GGGG}).\ \\ \ \\ 
Once the lower bound $\bar q_{min}$ is computed, one can come back to the $q$-dependent certification  condition (\ref{tgf56}) of Proposition \ref{enfinpropos} in order to compute for each $\bar q\ge \bar q_{min}$ the lower bound $\underline \varepsilon_0(\bar q)$ and the upper bounds $\bar \varepsilon_0(\bar q)$ of the admissible values of $\varepsilon_0$. 

\begin{figure}
\begin{center}
\input{res2.tex}
\end{center} 
\caption{Check of the certification feasibility for the chain of $n=2$ integrators. The condition (\ref{hg87650}) is satisfied by an interval of values of $\varepsilon_0$ including values satisfying (\ref{GGGG}). Successful values: $\bar q_{min}=0.18$, $\gamma_c=0.2$, $E_\mathbb{C}^1=0.3$.} \label{cas21} 
\end{figure}
\begin{figure}
\begin{center}
%
%
%
\definecolor{mycolor1}{rgb}{1,0,1}%
\begin{tikzpicture}

\begin{axis}[%
legend entries={$\underline\varepsilon_0$,$\bar\varepsilon_0$,$\varepsilon_0^{sol}$},
legend style={at={(0.9,0.75)},anchor=north east},
width=0.35\textwidth,
height=0.25\textheight,
scale only axis,
separate axis lines,
every outer x axis line/.append style={darkgray!60!black},
every x tick label/.append style={font=\color{darkgray!60!black}},
xmin=0.1,
xmax=15,
xlabel={$q(x)/\bar{q}_{min}$},
every outer y axis line/.append style={darkgray!60!black},
every y tick label/.append style={font=\color{darkgray!60!black}},
ymode=log,
ymin=1e-08,
ymax=0.1,
yminorticks=true,
title={$\underline{\varepsilon}_0(x)$, $\bar \varepsilon_0(x)$ and $\varepsilon_0^{sol}(x)$},
]
\addplot [
color=blue,
solid,
mark=o,
mark options={scale=0.5},
]
table[row sep=crcr]{
24.8327962737759 1e-08\\
20.8334015759057 1.38949549437314e-08\\
17.4739848522973 1.93069772888325e-08\\
14.6527537975843 2.68269579527973e-08\\
12.2840485117454 3.72759372031494e-08\\
10.2957964721207 5.17947467923121e-08\\
8.62736831236975 7.19685673001151e-08\\
7.22777155131376 1e-07\\
6.0933870619247 1.38949549437314e-07\\
5.13861548671092 1.93069772888325e-07\\
4.33434020282425 2.68269579527973e-07\\
3.65713678296091 3.72759372031494e-07\\
3.08726877281945 5.17947467923121e-07\\
2.60811960182267 7.19685673001153e-07\\
2.20571538811538 1e-06\\
1.86832553374275 1.38949549437314e-06\\
1.58613046550598 1.93069772888325e-06\\
1.35094801109779 2.68269579527973e-06\\
1.15601158806127 3.72759372031494e-06\\
1 5.17947467923121e-06\\
};
\addplot [
color=black,
solid,
mark=triangle,
mark options={scale=0.5},
]
table[row sep=crcr]{
1 5.17947467923121e-06\\
1.00271353075731 7.19685673001153e-06\\
1.00647182448768 1e-05\\
1.01167077897742 1.38949549437314e-05\\
1.01885066300236 1.93069772888325e-05\\
1.02874389481705 2.68269579527973e-05\\
1.04233462313383 3.72759372031494e-05\\
1.06092989543066 5.17947467923121e-05\\
1.08623961014885 7.19685673001153e-05\\
1.12045864622223 0.0001\\
1.16634038487329 0.000138949549437314\\
1.22724853962919 0.000193069772888325\\
1.30717720843501 0.000268269579527973\\
1.41073984449787 0.000372759372031494\\
1.54314441100096 0.000517947467923121\\
1.71018629089982 0.000719685673001151\\
1.91829299690118 0.001\\
2.17464276994686 0.00138949549437314\\
2.48736045301806 0.00193069772888325\\
2.86577996111667 0.00268269579527973\\
3.32075914988686 0.00372759372031494\\
3.86503803022087 0.00517947467923121\\
4.51363974402424 0.00719685673001151\\
5.28432138146609 0.01\\
6.19808712522176 0.0138949549437314\\
7.27977969964756 0.0193069772888325\\
8.55876860876594 0.0268269579527973\\
10.0697559830709 0.0372759372031494\\
11.8537235497355 0.0517947467923121\\
13.9590475904114 0.0719685673001151\\
16.442812924473 0.1\\
};
\addplot [
color=mycolor1,
solid,
line width=1pt,
]
table[row sep=crcr]{
1 3.10768480753873e-06\\
1.00271353075731 4.31811403800692e-06\\
1.00647182448768 6e-06\\
1.01167077897742 8.33697296623881e-06\\
1.01885066300236 1.15841863732995e-05\\
1.02874389481705 1.60961747716784e-05\\
1.04233462313383 2.23655623218896e-05\\
1.06092989543066 3.10768480753873e-05\\
1.08623961014885 4.31811403800692e-05\\
1.12045864622223 6e-05\\
1.16634038487329 8.33697296623883e-05\\
1.22724853962919 0.000115841863732995\\
1.30717720843501 0.000160961747716784\\
1.41073984449787 0.000223655623218896\\
1.54314441100096 0.000310768480753873\\
1.71018629089982 0.000431811403800691\\
1.91829299690118 0.0006\\
2.17464276994686 0.000833697296623883\\
2.48736045301806 0.00115841863732995\\
2.86577996111667 0.00160961747716784\\
3.32075914988686 0.00223655623218896\\
3.86503803022087 0.00310768480753873\\
4.51363974402424 0.00431811403800691\\
5.28432138146609 0.006\\
6.19808712522176 0.00833697296623883\\
7.27977969964756 0.0115841863732995\\
8.55876860876594 0.0160961747716784\\
10.0697559830709 0.0223655623218896\\
11.8537235497355 0.0310768480753873\\
13.9590475904114 0.0431811403800691\\
16.442812924473 0.06\\
};
\addplot [
color=red,
solid,
]
table[row sep=crcr]{
1 1e-08\\
1 0.1\\
};
\end{axis}
\end{tikzpicture}%
\end{center} 
\caption{Quadruple integrator: Evolution of the bounding values $\underline\varepsilon_0(q(x))$ and $\bar\varepsilon_0(q(x))$ and a possible state dependent precision $\varepsilon_0^{sol}(x)$ defined by (\ref{NBNBNB}).} \label{autreq4} 
\end{figure}
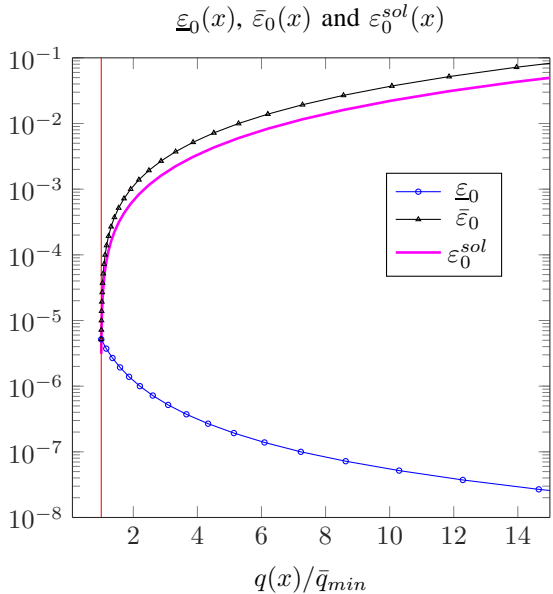

\ \\ \ \\ 
The state dependent sampling (\ref{sdsam}) can therefore be defined by the number of iteration associated to the precision $\varepsilon_0^{sol}$ given by:
\begin{eqnarray}
\varepsilon_0^{sol}(x):=(1-\lambda)\underline\varepsilon_0(q(x))+\lambda\bar\varepsilon_0(q(x)) \label{NBNBNB} 
\end{eqnarray}  
where $\lambda\in [0.5,0.9]$ in order to enhance high sampling period (higher values of $\varepsilon_0$) while keeping some security margin. \ \\ \ \\ 
Figures \ref{autreq4} shows the corresponding evolutions of the bounds $\underline \varepsilon_0(q)$ and $\bar\varepsilon_0(q)$ as functions of the ratio $q/\bar q_{min}$ for the quadruple integrator system ($n=4$) studied previously. This Figure clearly shows that when $q(x)$ is high, law precision (high values of $\varepsilon_0$) can be used. Although this is a known fact, the results proposed here gives a certified explicit computation of this feature. The Figure shows also clearly that attempt to achieve over-precise solution may lead to instability since there is a lower bounds on $\varepsilon_0$. 

\section{Conclusion}
In this paper a certification bound on the convergence of the fast gradient algorithm when applied to solve convex optimization problems with general inequality constraints with a prescribed level of sub-optimality is first given. The resulting bound is then used to derive a real-time implementation of MPC with state-dependent updating period leading to certified convergence of the resulting closed-loop to a neighborhood of the desired set-point. The proposed results clearly showed that the time needed to perform the elementary iteration is a key parameter in the resulting MPC implementation. To this respect, the proposed results can be used to afford limited computational power or to compute, for a given control problem and a given specification in terms of optimality and constraints fulfillment, the admissible computation power that need to be assigned. 

\appendix

\section{Appendix} \label{secappend} 
\subsection{Proof of Lemma \ref{lemmapropderive}} \label{prooflemmapropderive} 
This comes from the fact that $p^*$ is the unconstrained optimum of $f$ which means that:
\begin{eqnarray}
f_0(p^*)+\rho \psi(p^*)\le f_0(p_a)+\psi(p_a)=f_0(p_a)
\end{eqnarray} 
and since $\psi(p^*)\ge 0$, the last inequality gives $f_0(p^*)\le f_0(p_a)$. The inequality to be proved is therefore a simple consequence of the definition \ref{defdeD0Dpsi} of $D_0$. $\hfill \Box$
\subsection{Proof of Lemma \ref{propexpressphi}} \label{lemmapropexpressphi} 
Let us denote by $p_\psi$ the closest element of $\mathcal A_{\psi=0}$ to $p^*$. The triangular inequality implies:
\begin{eqnarray}
\|p-p_\psi\|&\le& \|p-p^*\|+\|p^*-p_\psi\|\le d(p)+\|p^*-p_\psi\|\nonumber \\ \label{triangular} 
\end{eqnarray} 
and because $\psi\in \mathcal F^1_{L_\psi}$:
\begin{eqnarray}
\psi(p)\le \psi(p_\psi)+\langle \psi^{'}(p_\psi),p-p_\psi\rangle+\dfrac{L_\psi}{2}\|p-p_\psi\|^2 \label{yh87} 
\end{eqnarray}
but since $p_\psi\in \mathcal A_{\psi=0}$, one has that $\psi(p_\psi)=0$ and $\psi^{'}(p_\psi)=0$ [because of the particular structure of the penalty], therefore (\ref{yh87}) becomes (because of (\ref{triangular})):
\begin{eqnarray}
\psi(p)\le \dfrac{L_\psi}{2}\|p-p_\psi\|^2\le \dfrac{L_\psi}{2}\Bigl[d(p)+\|p^*-p_\psi\|\Bigr]^2\label{AVVec} 
\end{eqnarray}  
It remains to prove that the term $\|p^*-p_\psi\|$ can be bounded so that the inequality (\ref{hgYYg6}) holds. 
Note that since $p^*$ minimizes $f$, one has:
\begin{eqnarray*}
f_0(p_\psi)+\rho\psi(p_\psi)\ge f_0(p^*)+\rho\psi(p^*)
\end{eqnarray*} 
and since $\psi(p_\psi)=0$ the last inequality leads to:
\begin{eqnarray}
\psi(p^*)&\le& \dfrac{1}{\rho}\bigl[f_0(p_\psi)-f_0(p^*)\bigr]\\
&\le& \dfrac{1}{\rho}\bigl[\|f^{'}_0(p^*)\|\cdot \|p^*-p_\psi\| +\dfrac{L_0}{2}\|p^*-p_\psi\|^2\bigr] \label{ghFFd} 
\end{eqnarray} 
Now let $p_u$ be the unconstrained minimizer of $f_0$, namely $f^{'}_0(p_u)=0$. Note that $p_u$ is uniquely defined since $\mu_0>0$ by assumption. Now by definition of $p^*$ and $p_u$, one has:
\begin{eqnarray}
f_0(p^*)+\rho\psi(p^*)\le f_0(p_u)+\rho\psi(p_u) \label{jjj8} 
\end{eqnarray} 
on the other hand, 
\begin{eqnarray}
f_0(p^*)\ge f_0(p_u)+\dfrac{\mu_0}{2}\|p^*-p_u\|^2 \label{jjj9} 
\end{eqnarray} 
By combining (\ref{jjj8})-(\ref{jjj9}), it comes that:
\begin{eqnarray*}
\|p^*-p_u\|\le \sqrt{\dfrac{2\rho}{\mu_0}\psi(p_u)}
\end{eqnarray*} 
This with the Lypschitz induced inequality gives:
\begin{eqnarray*}
\|f^{'}_0(p^*)-0\|\le L_0\|p^*-p_u\|\le L_0\sqrt{\dfrac{2\rho}{\mu_0}\psi(p_u)}=:\kappa_0^{'}\sqrt{\rho}
\end{eqnarray*} 
where $k_0^{'}:=L_0\sqrt{2\psi(p_u)/\mu_0}$.This last inequality together with (\ref{ghFFd}) implies:
\begin{eqnarray}
\psi(p^*)\le \dfrac{1}{\rho}\bigl[\kappa_0^{'}\sqrt{\rho}\|p^*-p_\psi\| +\dfrac{L_0}{2}\|p^*-p_\psi\|^2\bigr] \label{ghFFd12} 
\end{eqnarray} 
Now using (\ref{defdebeta}) in which $d(p^*,\mathcal A_{\psi=0})=\|p^*-p_\psi\|$ gives: 
\begin{eqnarray*}
\beta \|p^*-p_\psi\|^2\le \dfrac{1}{\rho}\bigl[\kappa_0^{'}\sqrt{\rho}\|p^*-p_\psi\| +\dfrac{L_0}{2}\|p^*-p_\psi\|^2\bigr]
\end{eqnarray*} 
and after straightforward manipulations, it comes that:
\begin{eqnarray}
\Bigl[\beta-\dfrac{L_0}{2\rho}\Bigr] \|p^*-p_\psi\| \le \dfrac{\kappa_0^{'}}{\sqrt{\rho}} \label{ghFFd25} 
\end{eqnarray} 
Now assuming that $\rho\ge L_0/\beta$, one obtains:
\begin{eqnarray*}
\|p^*-p_\psi\|\le \dfrac{2\kappa_0^{'}}{\beta\sqrt{\rho}}=\dfrac{2L_0\sqrt{2\psi_0(p_u)/\mu_0}}{\beta\sqrt{\rho}}
\end{eqnarray*} 
which together with (\ref{AVVec})  clearly ends the proof since the inequality (\ref{oju76756}) is a direct consequence of the fact that $d(p^*)=0$ by definition. $\hfill \Box$
\subsection{Proof of Lemma \ref{corcorcorhg6}} \label{proofcorcorcorhg6} 
Since $p_\psi$ is admissible and $p^{opt}$ is the optimal solution of the constrained problem, one necessarily has:
\begin{eqnarray}
f_0(p^{opt})\le f_0(p_\psi) \label{rfd54} 
\end{eqnarray} 
Moreover, since $f_0\in \mathcal F^1_{L_0}$, the following inequality holds:
\begin{eqnarray*}
f_0(p_\psi)\le f_0(p^*)+\|f_0^{'}(p^*)\|\cdot \|p^*-p_\psi\|+\dfrac{L_0}{2}\|p^*-p_\psi\|^2
\end{eqnarray*} 
and since $\|f_0^{'}(p^*)\|\le D_0$ (Lemma \ref{defdeD0Dpsi}):
\begin{eqnarray}
f_0(p_\psi)\le f_0(p^*)+D_0\|p^*-p_\psi\|+\dfrac{L_0}{2}\|p^*-p_\psi\|^2
\end{eqnarray} 
which together with (\ref{rfd54}) and (\ref{defdebeta}) of Assumption \ref{assphi} gives:
\begin{eqnarray}
\vert f_0(p^{opt})-f_0(p^*)\vert \le D_0\left[\dfrac{\psi(p^*)}{\beta}\right]^{\frac{1}{2}}+\dfrac{L_0}{2}\left[\dfrac{\psi(p^*)}{\beta}\right]
\end{eqnarray} 
This obviously ends the proof. $\hfill \Box$ 
\subsection{Proof of Lemma \ref{fgfYH}} \label{prooffgfYH} 
Assume that for some $p$ the following inequality hold:
\begin{eqnarray}
\vert f(p)-f(p^*)\vert \le \epsilon 
\end{eqnarray} 
this means that ($f\in \mathcal S^1_{\mu_0})$:
\begin{eqnarray}
\|p-p^*\|\le \left[\frac{2\epsilon}{\mu_0}\right]^{\frac{1}{2}} \label{gfty545875} 
\end{eqnarray} 
on the other hand:
\begin{eqnarray}
\vert f_0(p)-f_0(p^*)\vert \le D_0\|p-p^*\|+\dfrac{L_0}{2}\|p-p^*\|^2
\end{eqnarray} 
this together with (\ref{gfty545875}) gives the result. $\hfill \Box$ 
\subsection{Proof of Proposition \ref{mainresult}} \label{proofmainresult} 
{\sc Proof}. We shall first prove that when the algorithm stops, one has:
\begin{eqnarray}
\vert f(\hat p^*)-f(p^*)\vert \le \eta \label{whatever} 
\end{eqnarray}  
then we prove that when (\ref{whatever}) holds then $\hat p^*$ is an $\bar\varepsilon$-suboptimal solution of the original problem. To prove (\ref{whatever}), we shall distinguish two situations depending on the exit condition of step 10. Indeed, either $g(p_i)\le g_{min}:=\mu_0\sqrt{2\eta/L}$ in which case (\ref{whatever}) is satisfied since $f\in \mathcal S_{\mu_0,L}^1$. Or the algorithm stops after $\bar N(c,\gamma_0)$  iterations where $\gamma_0:=\eta\mu_0/[(L+\mu_0)f_0(p_0)]$ which implies (\ref{whatever}) by virtue of Corollary \ref{cortg6}. \ \\ \ \\
We shall now prove that when (\ref{whatever}) holds, one necessarily has:
\begin{eqnarray}
&&\vert f_0(\hat p^*)-f_0(p^{opt})\vert \le \varepsilon_0 \quad ;\quad \psi(\hat p^*)\le \varepsilon_\psi^2 \label{CCC} 
\end{eqnarray}  
\underline{Proof of $\psi(\hat p^*)\le \varepsilon_\psi^2$}\\ 
By the $\mu_0$-strong convexity of $f$, equation (\ref{whatever}) implies that $\|\hat p^*-p^*\|\le \sqrt{(2/\mu_0) \eta}$. Injecting this in (\ref{hgYYg6}) gives:
\begin{eqnarray*}
\psi(\hat p^*)\le \dfrac{L_\psi}{2}\left[\sqrt{\dfrac{2\eta}{\mu_0}}+\dfrac{\kappa_0}{\sqrt{\rho}}\right]^2
\end{eqnarray*}
So in order to prove that $\psi(\hat p^*)\le \varepsilon_\psi^2$, it is sufficient to prove the following two inequalities:
\begin{eqnarray*}
\sqrt{\dfrac{2\eta}{\mu_0}}\le \dfrac{\varepsilon_\psi}{2}\sqrt{\dfrac{2}{L_\psi}}\quad \mbox{\rm and}\quad \dfrac{\kappa_0}{\sqrt{\rho}}\le  \dfrac{\varepsilon_\psi}{2}\sqrt{\dfrac{2}{L_\psi}}
\end{eqnarray*}  
But the first inequality is satisfied because $\eta\le \eta_2$ while the second is satisfied because $\rho\ge \rho_1$.\ \\ \ \\ 
\underline{Proof of $\vert f(\hat p^*)-f_0(p^{opt})\vert\le \varepsilon_0$}\ \\ \ \\ 
Using the triangular inequality:
\begin{eqnarray*}
\vert f_0(\hat p^*)-f_0(p^{opt})\vert\le \vert f_0(\hat p^*)-f_0(p^*)\vert+ \nonumber \\ \vert f_0(p^*)-f_0(p^{opt})\vert
\end{eqnarray*} 
and using (\ref{onze}) and (\ref{jhGG034})  the last inequality gives:
\begin{eqnarray*}
&&\vert f_0(\hat p^*)-f_0(p^{opt})\vert\le D_0\left[\frac{2\eta}{\mu_0}\right]^{\frac{1}{2}}+\dfrac{L_0}{2}\left[\frac{2\eta}{\mu_0}\right]+\nonumber \\
&&D_0\left[\dfrac{\psi(p^*)}{\beta}\right]^{\frac{1}{2}}+\dfrac{L_0}{2}\left[\dfrac{\psi(p^*)}{\beta}\right]
\end{eqnarray*} 
therefore, the result can be obtained if the following inequality are satisfied:
\begin{eqnarray}
D_0\left[\frac{2\eta}{\mu_0}\right]^{\frac{1}{2}}+\dfrac{L_0}{2}\left[\frac{2\eta}{\mu_0}\right]&\le& \dfrac{\varepsilon_0}{2} \label{premierec}\\ 
D_0\left[\dfrac{\psi(p^*)}{\beta}\right]^{\frac{1}{2}}+\dfrac{L_0}{2}\left[\dfrac{\psi(p^*)}{\beta}\right]&\le &\dfrac{\varepsilon_0}{2}
\end{eqnarray} 
The first inequality is satisfied since $\eta\le \eta_1$ while the second is satisfied if:
\begin{eqnarray}
\left[\dfrac{\psi(p^*)}{\beta}\right]^{\frac{1}{2}}\le Z_1(\dfrac{\varepsilon_0}{2})
\end{eqnarray} 
But thanks to (\ref{oju76756}) [satisfied since $\rho\ge \rho_3$] this can be proved if the following inequality holds:
\begin{eqnarray}
\dfrac{L_\psi \kappa_0^2}{2\beta \rho}\le Z_1^2(\dfrac{\varepsilon_0}{2})
\end{eqnarray} 
which is satisfied because $\rho\ge\rho_2$.  $\hfill \Box$ 
\subsection{Proof of Proposition \ref{propD0}} \label{proofpropD0} 
\noindent Recall that in the specific case of QP problem, the definition of $D_0$ becomes
\begin{eqnarray*}
D_0:=\sup_{f_0(p)\le f_0(p_a)}\|Hp+F\|
\end{eqnarray*} 
But we have by assumption $\|p_a\|\le p_{max}$, which enables to write:
\begin{eqnarray*}
f_0(p_a)\le \dfrac{1}{2}\lambda_{max}(H)\left[\varrho(\mathbb P)\right]^2+\|F\|\cdot \varrho(\mathbb P)+\phi_0=:f
\end{eqnarray*} 
and since $f_0(p)\ge  \dfrac{1}{2}\lambda_{min}(H)\|p\|^2-\|F\|\|p\|+\phi_0$, the last inequality implies:
\begin{eqnarray*}
\|p\|&\le& \dfrac{\|F\|+\sqrt{\|F\|^2+2\lambda_{min}(H)\left[f-\phi_0\right]}}{\lambda_{min}(H)}=:\bar p
\end{eqnarray*} 
which obviously gives the results. $\hfill \Box$
\subsection{Proof of Lemma \ref{lemfondineq}} \label{prooflemfondineq} 
Using Assumption \ref{assdx} and \ref{assprederror} , it comes that:
\begin{eqnarray}
f_0(\hat p^*(t_{k+1}),x(t_{k+1}))\le f_0(\hat p^*(t_{k+1}),\hat x(t_{k+1}))+\nonumber \\ +K^0_\mathbb{C}\times\left[E^0_\mathbb{C}+E^1_\mathbb{C}\times \tau_k\right] \label{pr1} 
\end{eqnarray}  
Now by definition of $\tau_k$, the solution $\hat p^*(t_{k+1})$ satisfies
\begin{eqnarray*}
f_0(\hat p^*(t_{k+1}),\hat x(t_{k+1}))\le f_0(p^{opt}(t_{k+1}),\hat x(t_{k+1}))+\varepsilon_0^{(k+1)}
\end{eqnarray*}  
which together with Assumption \ref{assDelta} gives:
\begin{eqnarray}
f_0(\hat p^*(t_{k+1}),\hat x(t_{k+1}))\le&& f_0(p^{opt}(t_{k}),x(t_{k}))+\varepsilon_0^{(k+1)}\nonumber \\
&&-\Delta(\tau_k,x(t_k))\nonumber \\
&\le& f_0(\hat p^*(t_{k}),x(t_{k}))+\varepsilon_0^{(k)}\nonumber \\
&&+\varepsilon_0^{(k+1)}-\Delta(\tau_k,x(t_k))
\end{eqnarray} 
Using the last inequality in (\ref{pr1}) gives the result. $\hfill \Box$
\subsection{Proof of Lemma \ref{simplergloballemma}} \label{proofsimplergloballemma} 
By definition of (\ref{gfTTRTRDE}) of $\Delta$ and using (\ref{qx})  of Assumption (\ref{asslC}), it comes that:
\begin{eqnarray*}
\Delta(\tau,x)&\ge& \int_0^\tau \max\{0,q(x)-D_{\mathbb C}s\}ds\\
&=&\int_0^{\min\{\tau,q(x)/D_{\mathbb C}\}} (q(x)-D_{\mathbb C}s)ds\\
&=&\left[q(x)\tau-\dfrac{1}{2}D_\mathbb{C}\tau^2\right]_0^{\min\{\tau,q(x)/D_{\mathbb C}}
\end{eqnarray*}
which can be expressed using $\Gamma_\mathbb{C}(\tau,q)$ given by (\ref{gf52MLOK}).  $\hfill \Box$ 
\subsection{Proof of Proposition \ref{enfinpropos}} \label{proofenfinpropos} 
The first inequality in (\ref{jhghghjg}) together with Assumption \ref{assCphi} impliy that Corollary \ref{corcorcor} applies with $k=0$, $\mathbb C=\mathbb C_{\phi_0}$ and $\bar q:=q(x(t_k))$, therefore one has:
\begin{eqnarray}
&&f_0(\hat p^*(t_{1}),x(t_{1}))-f_0(\hat p^*(t_{0}),x(t_{0}))\le \nonumber \\
&&\varepsilon_0^{(0)}+R_{\tau_c}(\varepsilon_0^{(1)},\varepsilon_\psi,q(x(t_0)))\label{lhs45rewritten} 
\end{eqnarray} 
and since $\varepsilon_0^{(1)}=\varepsilon_0^{sol}(x(t_0))$, if  $q(x(t_0))> \bar q_{min}$ the inequality (\ref{tgf56}) gives:
\begin{eqnarray}
R_{\tau_c}(\varepsilon_0^{(1)},\varepsilon_\psi,q(x(t_0)))\le -\dfrac{\gamma_c\bar q_{min}^2}{3D_{\mathbb C_{\phi_0}}} \label{tgf56bis}
\end{eqnarray} 
and thanks to the second inequality in (\ref{jhghghjg}), the inequality (\ref{tgf56bis}) gives:
\begin{eqnarray}
\varepsilon_0^{(0)}+R_{\tau_c}(\varepsilon_0^{(1)},\varepsilon_\psi,q(x(t_0)))\le -\dfrac{\gamma_c\bar q_{min}^2}{6D_{\mathbb C_{\phi_0}}} \label{tgf56bisbis}
\end{eqnarray} 
This together with (\ref{lhs45rewritten}) implies that $f_0(\hat p^*(t_1),x(t_1))$ decreases meaning that the new pair is still in $\mathbb C_{\phi_0}$ and since $\varepsilon_0^{(1)}$ satisfies by assumption the second inequality in (\ref{jhghghjg}), the argumentation can be repeated to derive the properties of the next pair $(\hat p^*(t_2),x(t_2))$ meaning that the following inequality:
\begin{eqnarray*}
&&f_0(\hat p^*(t_{k+1}),x(t_{k+1}))-f_0(\hat p^*(t_{k}),x(t_{k}))\le -\dfrac{\gamma_c\bar q_{min}^2}{6D_{\mathbb C_{\phi_0}}}\label{lhs45rewrittenfin} 
\end{eqnarray*} 
is satisified as far as $q(x(t_k))$ remains grater than $\bar q_{min}$. This clearly implies that $x(t_k)$ converges to the limit set $\mathbb X_{min}$ defined by (\ref{defdeXmin}). \ \\ \ \\ 
regarding the constraints, note that the hard constraints are necessarily satisfied since they depend only on $p$ by assumption and that $\hat p^*(t_{k+1})$ satisfies by construction the hard constraints while allowing only for a violation of the soft constraints by an amount which is lower than $\varepsilon_\psi$, therefore, one has:
\begin{eqnarray}
c_i(\hat p^*(t_{k+1}),\hat x(t_{k+1}))\le \varepsilon_\psi\quad \forall i\in I_s
\end{eqnarray} 
which obviously gives (\ref{viuolatesfot}) by Assumptions \ref{assprederror} and \ref{assdx}. 
$\hfill \Box$
\subsection{Proof of Proposition \ref{proposphi0Cphi0}} \label{proofproposphi0Cphi0} 
Given $z_d$, one rewrite the cost function using the change of variable $y=z-z_d$ enables to write the cost function (\ref{costLMPC}) in the form
\begin{eqnarray*}
f_0(p,y)=\frac{1}{2} \begin{pmatrix}
p\crcr y
\end{pmatrix}^T \underbrace{\begin{pmatrix}
H&F_{11}\cr F_{11}^T&S_{11}
\end{pmatrix}}_{W_0} \begin{pmatrix}
p\cr y
\end{pmatrix}
\end{eqnarray*}  
which means that if $f_0(p,x)\le \phi$ then the following inequalities hold:
\begin{eqnarray}
\|p\|\le \left[\phi/\lambda_{min}(W_0)\right]^{\frac{1}{2}}\ ;\ \|y\|\le \left[\phi/\lambda_{min}(W_0)\right]^{\frac{1}{2}}
\end{eqnarray} 
The first inequality obviously gives (\ref{defdePbb}) while the second leads to:
\begin{eqnarray*}
\|z\|\le \|z_d\|+ \left[\phi/\lambda_{min}(W_0)\right]^{\frac{1}{2}}
\end{eqnarray*}  
which gives (\ref{gfGGvF}). 
 $\hfill \Box$
\subsection{Proof of Proposition \ref{propossssss76}} \label{proofpropossssss76} 
\noindent In order to prove that (\ref{expressionKC0}) satisfies (\ref{defdeKdep}), we use the definition of $f_0$ to write:
\begin{eqnarray*}
&&\|f_0(p,x_1)-f_0(p,x_2)\|=\nonumber \\ &&\left\|(F_1(x_1-x_2))^Tp+\|x_1\|_S^2-\|x_2\|_S^2\right\|\\
&&\le \|(F_1^Tp)^T(x_1-x_2)\|+2\lambda_{max}(S)\times \varrho(\mathbb X)\times \|x_1-x_2\|\\
&&\le \left[\|F_1^T\|\times \varrho(\mathbb P)+2\lambda_{max}(S)\times \varrho(\mathbb X)\right]\cdot \|x_1-x_2\|
\end{eqnarray*} 
which proves (\ref{expressionKC0}). \ \\ \ \\  It remains to prove that $K_{\mathbb C}^1$ defined by (\ref{expressionKC1}) satisfies (\ref{defdeKdepsi}) we first note that:
\begin{eqnarray*}
\psi(x):=\sum_{i=1}^{n_c}\left[r_i(p,x)\right]^2
\end{eqnarray*} 
with $r_i(p,x)=\max\bigl\{0,A_ip-B_i^0-B_i^1x\bigr\}$. Therefore:
\begin{eqnarray*}
\|\dfrac{\partial \psi}{\partial x}\|\le 2\sum_{i=1}^{n_c}\vert r_i(p,x)\vert \cdot \|B_i^{(1)}\|
\end{eqnarray*} 
and using the inequalities expressing the equivalence of the $L_2$ and $L_1$ norms, the last inequality gives:
\begin{eqnarray*}
\|\dfrac{\partial \psi}{\partial x}\|&\le& 2n_c\sum_{i=1}^{n_c}\vert r_i(p,x)\vert^2 \cdot \|(B^{(1)})^T\|\\ 
&\le& 2n_c\left[\psi(p,x)\right]\times \|(B^{(1)})^T\|\\ 
&\le& 2n_c\times \psi^{max}\times \|(B^{(1))^T}\|
\end{eqnarray*} 
which obviously gives (\ref{expressionKC1}). $\hfill \Box$
\bibliographystyle{plain}
\bibliography{fast_gradient}

\begin{thebibliography}{10}

\bibitem{IJCobs:1999}
M.~Alamir.
\newblock Optimization based nonlinear observers revisited.
\newblock {\em International Journal of Control}, 72(13):1204--1217, 1999.

\bibitem{alamirecc13}
M.~Alamir.
\newblock Monitoring control updating period in fast gradient-based {NMPC}.
\newblock In {\em Proceedings of the European Control Conference (ECC2013)},
  Zurich, Switzerland, 2013.

\bibitem{alamirecc14}
M.~Alamir.
\newblock Fast {MPC}, a reality steered paradigm: Key properties of fast {NMPC}
  algorithms.
\newblock In {\em Proceedings of the European Control Conference (ECC2014)},
  Strasbourg, France, 2014.

\bibitem{Biegler:2010}
L.~T. Biegler.
\newblock {\em Nonlinear Programming}.
\newblock Society for Industrial and Applied Mathematics, 2010.

\bibitem{Bemporad2012}
A.~Bomporad and P.~Patrinos.
\newblock Simple and certifiable quadratic programming algorithms for embedded
  linear model predictive control.
\newblock In {\em Proceeding of the IFAC Nonlinear Predictive Control
  Conference}, Noordwijkerhout, NL, 2012.

\bibitem{Bonne:2014arxiv}
F.~Bonne, M.~Alamir, and P.~Bonnay.
\newblock Experimental investigation of control updating period monitoring in
  industrial {PLC}-based fast {MPC:} application to the constrained control of
  a cryogenic refrigerator. arxiv:1406.6281.
\newblock 2012.

\bibitem{Domahidi2012o}
A.~Domahidi, A.~U. Zgraggen, M.~N. Zeilinger, and M.~Morariand~C. Jones.
\newblock Efficient interior point methods for multistage problems arising in
  receding horizon control.
\newblock In {\em Proceeding of the IEEE Conference on Decision and Control},
  Hawaii, USA, 2012.

\bibitem{Ferreau:2008}
H.~J. Ferreau, H.~G. Bock, and M.~Diehl.
\newblock An on line active set strategy to overcome the limitations of
  explicit mpc.
\newblock {\em Int. J. Robust Nonlinear Control}, 18:816--830, 2008.

\bibitem{Jones2012}
C.~N. Jones, A.~Domahidi, M.~Morari, S.~Richter, F.~Ullmann, and M.~Zeilinger.
\newblock Fast predictive control: Real-time computation and certification.
\newblock In {\em Proceeding of the IFAC Nonlinear Predictive Control
  Conference}, Noordwijkerhout, NL, 2012.

\bibitem{Mayne:2000}
D.~Q. Mayne, J.~B. Rawlings, C.~V. Rao, and P.~O. Scokaert.
\newblock Constrained model predictive control: Stability and optimality.
\newblock 36:789--814, 2000.

\bibitem{McGovern:2000}
L.~K. McGovern.
\newblock Computationak abalysis of real-time convex optimization for control
  systems. {Ph.D}. {D}issertation, {M}assachusetts {I}nstitute of {T}echnology.
  {D}epartment of {A}eronautics and {A}stronautics, {C}ambridge, {MA, USA}.
\newblock 2000.

\bibitem{Nesterov1983}
Y.~Nesterov.
\newblock A method of solving a convex programming problem with convergence
  rate {O}(1/k$^2$).
\newblock {\em Soviet Mathematics Doklady}, 27(2):372--376, 1983.

\bibitem{Nesterov2004}
Y.~Nesterov.
\newblock {\em Introductory lectures in convex optimization: a basic course}.
\newblock Kluwer Academic Publishers, 2004.

\bibitem{Richter:2012}
S.~Richter, C.N. Jones, and M.~Morari.
\newblock Computational complexity certification for real-time mpc with input
  constraints based on the fast gradient method.
\newblock {\em Automatic Control, IEEE Transactions on}, 57(6):1391--1403, June
  2012.

\bibitem{Zavala:2008}
V.~M. Zavala, C.~D. Laird, and L.~T. Biegler.
\newblock Interior point decomposition approaches for parallel solution of
  large scale nonlinear parameter estimation problems.
\newblock {\em Chemical Engineering Science}, 63(19):4834--4845, 2008.

\bibitem{Zhiwen:2000}
Zhiwen Zhu and H.~Leung.
\newblock Adaptive identification of nonlinear systems with application to
  chaotic communications.
\newblock {\em Circuits and Systems I: Fundamental Theory and Applications,
  IEEE Transactions on}, 47(7):1072--1080, Jul 2000.

\end{thebibliography}

\ifCLASSOPTIONcaptionsoff
  \newpage
\fi

\end{document}